\documentclass[openacc]{rsproca_new}

\usepackage{graphicx}%
\usepackage{amsmath,amssymb,amsfonts}%
\usepackage[square,sort,comma,numbers]{natbib}
\usepackage{subcaption} 
\usepackage{bm}
\usepackage{bbm}
\usepackage[T1]{fontenc}
\usepackage{lmodern} 
\usepackage{microtype} 

\def\rev#1{{{\textcolor{black}{ #1}}}} 

\def\add{}

\titlehead{Research}

\begin{document}

\title{Using spatial extreme-value theory with machine learning to
model and understand spatially compounding weather extremes}

\author{
Jonathan Koh$^{1,2}$, Daniel Steinfeld$^{2,3}$ and Olivia Martius$^{2}$}

\address{$^{1}$Seminar for Statistics, ETH Z\"urich\\ 
$^{2}$Institute of Geography, Oeschger Centre for Climate Change Research, University of Bern \\
$^{3}$GVZ Geb\"audeversicherung Kanton Z\"urich}

\subject{Artificial Intelligence, Atmospheric science, Statistics}



\corres{Jonathan Koh\\
\email{jonathan.koh@stat.math.ethz.ch}}

\begin{abstract}
When extreme weather events affect large areas, their regional to sub-continental spatial scale is important for their impacts. We propose a novel machine learning (ML) framework that integrates spatial extreme-value theory to model weather extremes and to quantify probabilities associated with the occurrence, intensity, and spatial extent of these events. Our approach employs new loss functions adapted to extreme values, enabling our model to prioritize the tail rather than the bulk of the data distribution. Applied to a case study of Western European summertime heat extremes, we use daily 500-hPa geopotential height fields and local soil moisture as predictors to capture the complex interplay between local and remote physical processes. Our generative model reveals that different facets of heat extremes \add{are influenced by individual circulation features, such as the relative position of upper-level ridges and troughs that are part of a large-scale wave pattern. This enriches our} process understanding from a data-driven perspective. Our approach can extrapolate beyond the range of the data to make risk-related probabilistic statements. It applies more generally to other weather extremes and offers an alternative to traditional physical and ML-based techniques that focus less on the extremal aspects of weather data. \absbreak 
\end{abstract}

\rsbreak


\section{Introduction}\label{sec1}

Extreme weather events profoundly impact society and ecosystems, with their impacts being exacerbated when such events are temporally or spatially compounding \citep[e.g., ][]{kornhuber2020, zscheischler2017, zscheischler2020}. Extreme heat events are examples of such high-impact weather events as they affect crop yields, wildfire occurrence, energy demand, transportation systems, human mortality and more \citep[e.g.,][]{coumou2012, vogel2019, vicedo-cabrera2021, white2023}. Notably, the record-breaking impacts of the European heatwaves of 2003, 2010, 2018, and 2022 resulted from events of record-shattering intensity and large spatial extent \citep{schar2004, matsueda2011, barriopedro2011, russo2015, drouard2019, rousi2022a, mitchell2019, lhotka2022, faranda2023}. \add{Spatially compounding events refer to (extreme) weather or climate events that affect multiple locations simultaneously, which thereby compound the impacts of the events \citep{zscheischler2020}}.

The study of these record-breaking extreme weather events has traditionally relied on physics-based numerical models. Studies utilise reforecast archives \citep{Kelder.2020.UNSEEN}, \add{ensemble boosting} \citep{Gessner.2021.EnsembleRe, Fischer.2023} or weather simulators \citep{cadiou2024_sub} to empirically quantify the return period of an event. While these models are foundational for current meteorological predictions, computational costs still limit the number of weather events that can be simulated, making it challenging to assess the probabilities of events with very extreme intensity and / or spatial extent in the current climate, let alone in a changing one.

Concurrently, rapid progress has been made in the quality of machine learning (ML)-based weather forecasts, such as with the introduction of FourCastNet \citep{pathak2022fourcastnet}, Pangu-Weather \citep{Pangu.2023.Nature} and Graphcast \citep{lam.graphcast.2023}. These models have much lower computational costs and can, in principle, simulate thousands of years of weather evolution quickly. However, these methods, along with approaches that model spatially compounding events using Gaussian processes directly \citep[e.g.,][]{mascolo.2024}, are trained by minimising errors based on the root mean squared error, which inherently assumes the target 
of interest is the conditional mean. 
This may not be suitable when the focus is on extreme values, and can lead to unsuitable penalisation of extreme forecasts and inaccurate predictions of their spatial extent \citep{ECMWF_blog_2023}. 


Our methodology proposes an alternative to the aforementioned approaches, by only using data from the distributional tail to fit ML models that optimise new loss functions motivated by extreme-value theory (EVT); this theory provides an elegant and mathematically justified framework to estimate the probabilities of rare events by extrapolating beyond the range of the data. Modelling spatially dependent weather extremes is particularly challenging, and \add{we appeal to the} sub-field of spatial EVT \citep{davison2012, Huser.Wadsworth.2022} to address this difficulty. 

ML tools have recently been combined with univariate EVT to predict conditional distributions above a high threshold, using gradient boosting \citep{Koh.2021b, velthoen.2021}, random forest \citep{Gnecco.2022}, and neural network architectures \citep{Richards.2022, Pasche.2022}. These studies rely on models like the Generalized Pareto distribution (GPD), but \add{usually} impose stringent spatial independence assumptions \add{in situations with many predictors}. \add{Furthermore, those that utilise many predictors to model variation in the spatial dependence of extreme events using EVT suffer from computational bottlenecks in statistical settings \citep[e.g.,][]{Koh.2024}. The situation is alleviated using ML methods, e.g., \cite{majumder.2023} utilise predictors in their spatial extremes applications using ML models, though not in a peak-over-threshold setting as we do here, while \cite{zhang2023.flexible} incorporate non-stationarity into their model via variational autoencoders without utilising predictors.

Analogous to the role of the GPD in univariate EVT, generalised $r$-Pareto processes \citep{deFondeville.2022} are suitable models for spatial extremes because they provide asymptotic approximations to the distribution of functional exceedances of appropriately re-scaled random fields. Parametric forms 
of these processes have been developed, most notably the Brown--Resnick \cite{brown1977extreme, engelke.2020} model, which is based on Gaussian processes. These processes are flexible enough to capture complex spatial dependencies. 

The predictive power of our proposed model is rooted in its ability to utilize predictors based on physical and dynamical reasoning. For example, multiple physical processes are involved in the formation of heat extremes in the midlatitudes \citep[e.g.,][]{rothlisberger2023, barriopedro2023, Noyelle2024}. On synoptic timescales, persistent large-scale anticyclones embedded in the midlatitude jet stream known as atmospheric blocks 
affect surface temperatures through horizontal temperature advection at their flanks and adiabatic warming due to atmospheric subsidence and clear-sky radiative forcing 
in their center \citep{pfahl2012, sousa2017, kautz2022, suarez-gutierrez2020, domeisen2023}. Beneath the anticyclones, land–atmosphere feedbacks can further enhance temperatures, with low soil moisture limiting evaporative cooling \citep[e.g.,][]{fischer2007, seneviratne2006, seneviratne2010, sousa2020, zhang2023, tuel2023}. 
Figure \ref{fig:intro_2022} illustrates the atmospheric conditions during the European summer heatwave in July 2022, when temperatures rose above 40$^{\circ}$C (more than 4 standard deviations away from the 1991--2020 mean climate) in UK for the first time. The large-scale circulation was characterised by a strongly meandering upper-level jet stream around a Rossby wave pattern, featuring a trough (negative Z500 anomaly) over the eastern Atlantic Ocean and an intense anticyclone (positive Z500 anomaly) over \rev{Europe}. Underneath the anticyclone, a negative soil moisture anomaly was co-located with the heat extreme. 

\begin{figure*}[t]
    \includegraphics[width=.99\textwidth]{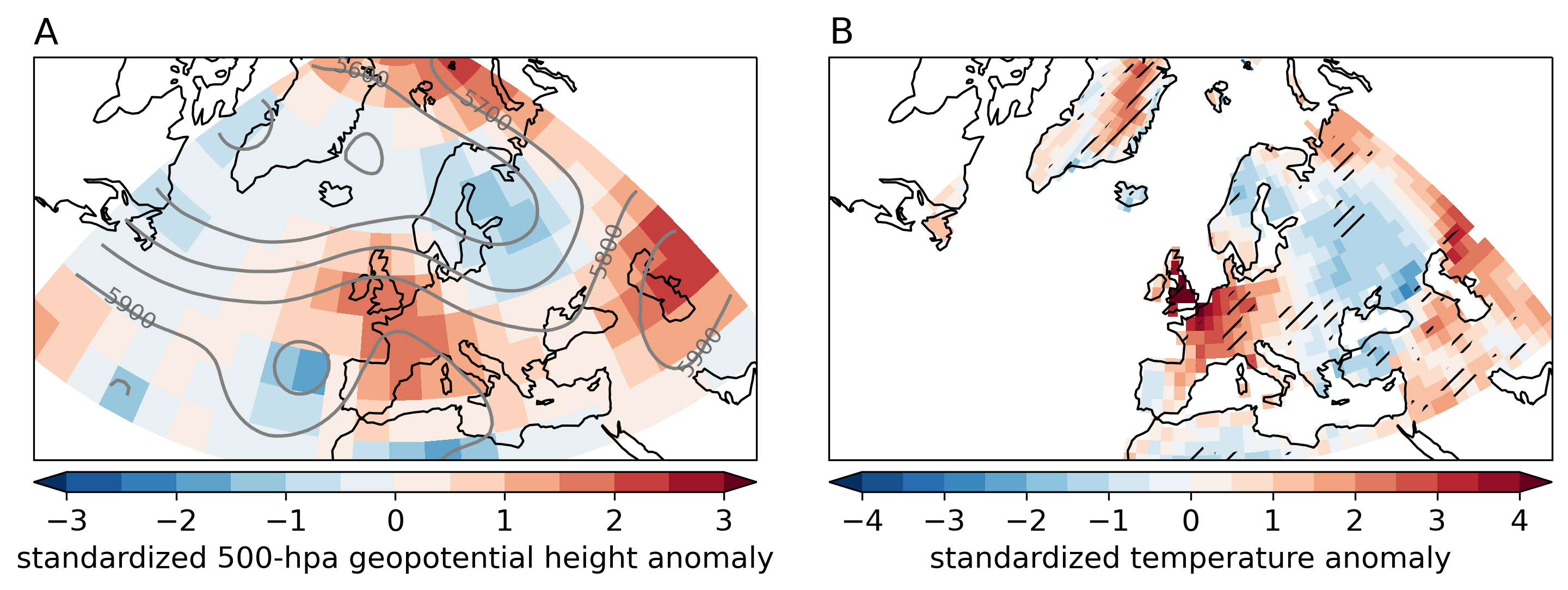}
    \caption{Characteristics of the 2022 European heatwave on July 19 from the ECMWF ERA5 reanalysis: (A) Standardised anomalies of 500-hPa geopotential height (Z500, shading interval -3 and 3) and absolute 500-hPa geopotential height contours (Z500$^\prime$, gray contour interval 5400m and 5800m every 100m), (B) standardised anomalies of surface 2m temperature (T2M, shading interval -4 and 4) and standardised anomalies of soil moisture (SM, hatching for areas where the SM is below minus one standard deviation). Calculation of standard deviations and anomalies is described in Section~2.}    
   \label{fig:intro_2022}
\end{figure*}

The relevance of the above-mentioned processes varies across heat extremes \citep{rothlisberger2023}, and they may interact in complex ways \citep{miralles2014, wehrli2019, xu2021}. 
The lack of physical understanding of these complex interactions during extreme-causing weather patterns impedes accurate prediction of heat extreme intensity and spatial extent \citep{horton2016, domeisen2023, barriopedro2023}. Our methodology seeks to address this knowledge gap. 

We develop three ML models 
that predict the occurrence probability (occurrence sub-model), the local intensity at each grid point (intensity sub-model), and the spatial extent (spatial dependence sub-model) of high-exceedance/extreme temperature events, by fitting a Brown--Resnick $r$-Pareto process to our data. The proposed models quantify temporal non-stationarity in the spatial extremal dependencies, while also being the first attempt in literature to do so with the incorporation of many predictors using EVT. In the case study of heat extremes over Western Europe showcasing our methodology, the predictors are daily fields of atmospheric circulation patterns over the Euro-Atlantic region and local or regional soil moisture summaries. 
Inspection of variable importance in our fitted models thus enriches our process understanding of summertime heat extremes from a data-driven perspective.

\section{Preliminaries}\label{sec:methods}

\subsection{Data}\label{sec:data}

We use daily gridded reanalysis data from the ECMWF-ERA5 reanalysis \citep{era5.2020} in the Atlantic–European region (60$^{\circ}$W--60$^{\circ}$E, 30$^{\circ}$N--90$^{\circ}$N) for summer (June to August) over the period 1959--2022. In a reanalysis, observations are assimilated into a numerical weather prediction model to produce the best estimate of the state of the atmosphere. \add{Throughout the paper, we superscript weather variables with an apostrophe to represent their raw versions.} We use the years 1959--2017 for training and 2018--2022 for testing the model performance. From the reanalysis, we use daily means of 2m land surface air temperature (T2M$^\prime$), 500-hpa geopotential height (Z500$^\prime$) and soil moisture integrated over a depth of 28\,cm (SM$^\prime$) at a horizontal longitude-latitude resolution of 1.875 degree for T2M$^\prime$ and SM$^\prime$, and 5.625 degree for Z500$^\prime$. \add{Our grid resolution choice is linked to the domain selection and is guided by expert knowledge; in our case, the domain and grid resolution was chosen to capture the typical size of the weather systems associated with heat waves, such as high- and low-pressure systems. We consider our study here as a proof of concept, and in theory our approach could be applied to any grid-resolution, which should be guided by the specific application and required spatial details.} The ERA5 data can be downloaded freely from the Copernicus Climate Data Store\footnote{\href{https://climate.copernicus.eu/climate-reanalysis}{https://climate.copernicus.eu/climate--reanalysis}}. For spatial modelling purposes, we project the \add{original longitude-latitude coordinates using the Universal Transverse Mercator projection. Thus, the distances used are great circle distances (in 100 km).}

The following preprocessing steps are applied to the 
response T2M$^\prime$ and predictors (Z500$^\prime$ and SM$^\prime$). As we do not consider climate change effects in our case study, the fields are first detrended by removing the linear trend for each grid point over the analysed baseline period (1959--2022). We then compute standardised anomalies by removing the climatological mean and dividing by its standard deviation (Figure~\ref{fig:t2m_anom}A,B). The climatological mean and standard deviation are defined as the 31-day running mean and standard deviation, centered around each date of the year and averaged over the World Meteorological Organization-defined reference period 1991--2020.

\begin{figure}[t]
\centering
    \includegraphics[width=0.7\textwidth]{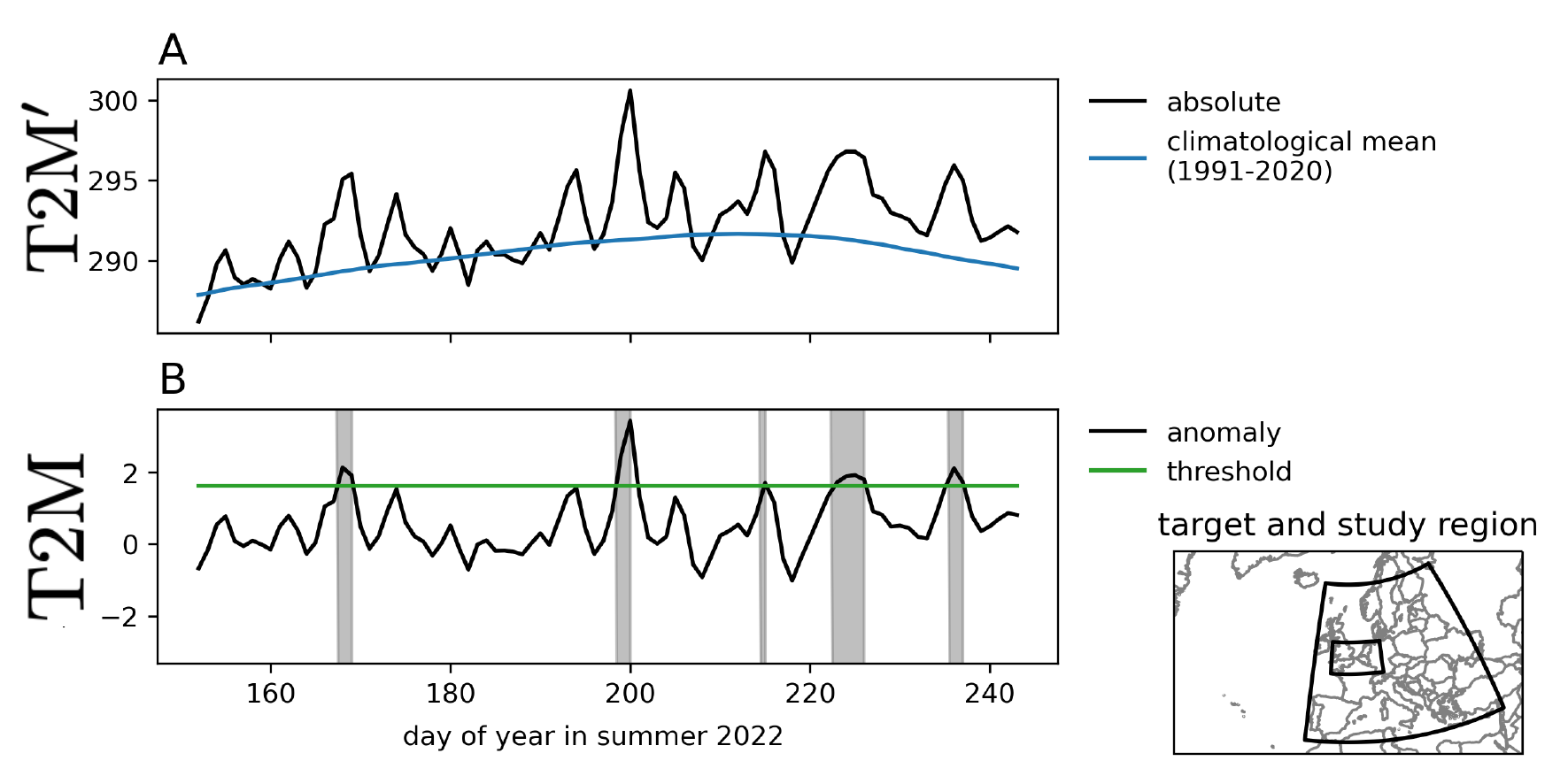}
   \caption{\add{(A) Temporal evolution of absolute value in summer 2022 (black) and climatological mean averaged over the target region (blue, 1991--2020) for T2M$^\prime$ (K). (B) Temporal evolution of the daily standardised anomalies (T2M) in summer 2022 averaged over the target region (black, with respect to climatological mean) and 95th percentile threshold (green) used for the definition of hot extremes (grey shaded) in the target region. The target and study region are shown on the map in the bottom right. Further descriptions and postprocessing steps involved in obtaining T2M and T2M$^\prime$ (K) are given in Section~\ref{sec:data}.} }   
   \label{fig:t2m_anom}
\end{figure}

The 500-hpa geopotential height predictor is further temporally averaged over the previous five days to remove higher frequency components, i.e., transient Rossby waves, and to isolate persistent larger-scale circulation patterns. 
Similarly, the soil moisture predictor is temporally averaged over the previous 15-day period. We call the corresponding postprocessed response T2M, and the predictors Z500 and SM. 


\subsection{Spatial exceedances}\label{sec:grp}

Let $S$ denote our study region, a compact subset of $\mathbb{R}^2$, and let $\mathcal{F}$ be the space of real-valued continuous functions on $S$. \add{Spatial extreme events of} the T2M random field $Y=\{Y(\bm s): \bm s\in S \} $ can be characterised \add{using a peaks-over-threshold approach based on} $r$-exceedances, defined as events $\{r(Y) \geq u\}$ for some threshold $u\geq 0$ and $r$ a continuous mapping from $\mathcal{F}$ to $\mathbb{R}$. \add{The function $r$ is commonly referred to as a risk functional, whose choices depend on the application. For instance, \cite{defondeville.2018} considered $r(Y)=\left(\int_{\bm s\in S} \mid Y(\bm s)\vert^p \mathrm{d}\bm s \right)^{1/p}$, $p>0$ in their case study on extreme rainfall; the value of the risk functional approaches $\max_{\bm s\in S} \vert Y(\bm s) \vert $ as $p$ increases, while the contribution of each location to the value increases with decreasing $p$. In this work, we restrict our focus to a linear risk functional which satisfy $r(x+y) = r(x) + r(y)$, for any $x,y\in\mathcal{F}$.}

We assume that $Y$ is generalised regularly varying with constant \add{tail index $\xi\in\mathbb{R}$ characterising the marginal tail decay rate, which, along with technical assumptions on $Y$ \citep[detailed in Theorem 2 of][]{deFondeville.2022}}, ensure that the tail limit processes for rescaled $Y$ conditional on an increasingly large $r$-exceedances of \add{$Y$}, converges to a generalized $r$-Pareto process (GrP), i.e.,
\begin{equation}\label{eq:rpareto.convergence}
    \operatorname{Pr}\left[ \frac{Y-b_{n}}{r\left(a_{n}\right)} \in \cdot \Bigl\vert r\left({Y}\right) \geqslant u_n\right] \rightarrow \operatorname{Pr}(P \in \cdot), \quad n \rightarrow \infty,
\end{equation}
where $\left\{a_{n}\right\}_{n=1}^{\infty}>0$ and $\left\{b_{n}\right\}_{n=1}^{\infty}$ are sequences of continuous functions, $u_{n}=r(b_n)$ 
is an increasing sequence, and $P$ is a GrP \add{with scale function $A$. Here we assume $a_n\approx r(a_n) A(\bm s)$, $\bm s\in \mathcal{S}$, for $n$ large, which holds in many applications where the marginal distributions of $Y$ belong to a location-scale family \citep{deFondeville.2022}.} 
In our case study, we assume that $u_n$ being the $95^\text{th}$ percentile of the empirical realisations of $r(Y)$ is sufficiently high for \eqref{eq:rpareto.convergence} to hold, and this also lies within the range over which the estimated tail index for the realisations of $r(Y)$ is stable. Relaxing the assumption of a \add{constant $\xi$} is possible: this parameter can be allowed to vary, 
and then observations can be transformed to have a common marginal distribution, such as the unit Fréchet or unit Pareto, so exceedances may then be defined on this transformed scale. However, a common tail decay restriction is needed to define the exceedances directly on the original process as we do here, as otherwise the grid point with the heaviest tail would dominate the limit distribution, leading to unrealistic models; \cite{deFondeville.2022} discuss this in more detail.

\subsection{GrP based on Gaussianity}\label{sec:br}
Equation \eqref{eq:rpareto.convergence} shows that GrPs appear as theoretically motivated limiting models for $r$-exceedances, akin to the Generalized Pareto distribution (GPD) for univariate exceedances \citep{Davison-Smith.1990}. \add{A GrP with scale function $A$} can be represented as \citep{Dombry-Ribatet.2015}:
$$
P = \add{A}/\xi \left[ \left\{R W/ \add{r\left(W^\xi\right)^{1/\xi} }\right\}^\xi - 1\right] ,
$$
where \add{the convention $ \{ z^\xi - 1 \}/\xi = \log(z)$, $z>0$, holds when $\xi=0$.} The scalar $R$ is unit Pareto and independent of the stochastic process $W$ satisfying $E\{ W(\bm s)\} = 1$, for all $\bm s\in \mathcal{S}$ and taking values in the unit $L_1$-sphere $\left\{y \in \mathcal{F}_{+}:\|y\|_{1}=1\right\}$, where $\mathcal{F}_+$ \add{is the subset of $\mathcal{F}$ containing only non-negative functions that are not everywhere zero} and $\| \cdot \|_{1}$ is the 1-norm on $\mathcal{F}_+.$ 


A flexible class of models 
is the Brown-Resnick \cite{brown1977extreme, engelke.2020} \rev{process}, where $W$ is a log-Gaussian \add{random function} whose \add{dependence is determined by a} Gaussian process that is centered and intrinsically stationary. 
These models are attractive to furnish models for extremal dependence as they are completely characterised by the \add{semivariogram} of the \add{underlying} Gaussian process $\gamma(\bm s_1, \bm s_2)=\add{\dfrac{1}{2}} \operatorname{var}\left\{W\left(\bm s_1\right)-W(\bm s_2)\right\}$, $\bm s_1,\bm s_2 \in S$. In our study we use the 
powered semivariogram 
\begin{align}\label{eq:covariance}
    \gamma(\bm s_1, \bm s_2) = \left\{\dfrac{\left|\left| \bm s_1- \bm s_2 \right| \right|}{\exp(\theta^\text{extent})} \right\}^\alpha,
\end{align}
where $\alpha\in(0,2]$ is known as the smoothness parameter. For a fixed $\alpha$, \rev{the parameter $\theta^\text{extent}\in\mathbb{R}$} can be interpreted in terms of the characteristic size of individual $r$-exceedance events. \rev{A simple form of geometric anisotropy without rotation can be introduced to the model, such that, for example, the coordinates in the $y$-dimension could be multiplied by $\exp(\theta^\text{scale})$, where $\theta^\text{scale} \in \mathbb{R}$.} 
Jointly, the two parameters $\theta^\text{extent}$ and $\theta^\text{scale}$ would then control the relative spatial extent in the east-west compared to the north-west direction. 

This framework models the spatial extremal dependence of the entire temperature field. More succinctly, this dependence can be summarised in a pairwise fashion for grid points $\bm s_1, \bm s_2 \in \mathcal{S}$ as 
\begin{align}\label{eq:condprob}
\pi_{u_q}(\bm s_1, \bm s_2)= \operatorname{Pr}[&Y\left(\bm s_2\right)>u_{q}\left(\bm s_2\right) \vert 
\left\{Y(\bm s_1)>u_{q}(\bm s_1)\right\} \cap\{r(Y) \geqslant u\}],
\end{align}
and approaches $2\left(1-\Phi\left[\{\gamma(\bm s_1, \bm s_2) / 2\}^{1 / 2}\right]\right)$ when $q\rightarrow 1$ for the Brown-Resnick model, where $u_{q}(\bm s)$ denotes the $q$ quantile of $Y(\bm s)$, $u\geq 0$ is the chosen threshold, and $\Phi$ denotes the standard normal cumulative distribution function. \add{For a fixed $u$ not depending on $q$, the conditioning event $\{r(Y)\geq u\}$ has no theoretical impact on the convergence of \eqref{eq:condprob} to the limit as $q\rightarrow 1$. However, in practice the choice of the risk functional $r$ and finite threshold $u$ may impact this convergence.} Given an $r$-exceedance and that the response at a grid point $\bm s_1$ exceeds a high threshold, \eqref{eq:condprob} evaluates the probability that another grid point $\bm s_2$ also exceeds a high threshold.

\subsection{Distributional tree boosting}\label{sec:boosting}

The development and dissemination of open-source packages such as \texttt{gbm} \citep{greenwell.gbm} and \texttt{xgboost} \citep{Chen.2016} have partly fueled the popularity of gradient tree boosting \citep{friedman.2001} methods over the last decade. When appropriately tuned and/or used in an ensemble with other ML approaches, these methods can give extremely accurate predictions in various classification and regression tasks \citep{bojer.2021}.

Let $\{(\bm x_t, \bm y_t) \}$, $\bm x_t \in \mathbb{R}^p $, $\bm y_t \in \mathbb{R}^D$, $t=1,\dots,T$, be a dataset with $T$ pairs of a response with dimension $D\in\mathbb{N}$, and $p\in\mathbb{N}$ predictors. A gradient tree boosting model gives an estimate $\hat{\theta}$ that is a sum of regression trees \citep{Breiman.1984}, i.e.,
\begin{equation}\label{eq:boosting_estimate}
    \hat{\theta}_t = \sum_{i=1}^{N} f_i(\bm x_t) , \quad f_i \in \rev{\mathcal{B}},
\end{equation}
where \rev{$\mathcal{B}$} is the space of regression trees. In the parametric distributional boosting framework, $\hat{\theta}_t$ represents a predicted parameter of a given model for the conditional distribution of the response given the predictors at time $t$. \add{For instance, $\hat{\theta}_t$ could represent a transformation of the predicted probability of occurrence of an $r$-exceedance in day $t$, or a predicted measure of the spatial extent of $r$-exceedances, such as $\theta^\mathrm{extent}$ in \eqref{eq:covariance}.} Other, assumption-free distributional boosting approaches have also been proposed \citep[e.g., ][]{Friedman.pnas.2020}, though we do not tackle these here.  


The set of trees used in \eqref{eq:boosting_estimate} are usually learnt by minimizing a regularised objective function in a greedy iterative fashion; we add the tree that minimizes an objective function $\mathcal{O}$ at each iteration with a stepsize hyperparameter dictating the regularisation. More precisely, let \add{$\hat{\theta}_t^{(j)}$} be the boosting estimate for the $t$-th observation at the $j$-th iteration, \add{where $j\in \{1,\dots, N$\}}. Many variants of this algorithm \add{exist}, but a popular one \cite{Friedman.etal.2000} uses a second-order approximation for the objective, so a tree $f_j$ is added at each iteration to minimize 
\begin{equation}\label{eq:objective}
 \sum_{t=1}^T {\left\{ \mathcal{L}(\bm y_t, \hat{\theta}_t^{(j-1)} ) + g_t f_j(\bm x_t) + \add{ \dfrac{1}{2} }h_t f_j^2(\bm x_t) \right\}} + \Omega(f_j), 
\end{equation}
where $\mathcal{L}$ is a differentiable loss function, $\Omega$ is a measure of complexity of the tree which prevents overfitting, $g_t= \partial \mathcal{L}(y_t, z_t )/ \partial z_t \vert_{z_t=\hat{\theta}_t^{(j-1)}}$  and $h_t= \partial^2 \mathcal{L}(y_t, z_t )/ \partial z_t^2 \vert_{z_t=\hat{\theta}_t^{(j-1)}}$. \add{The gradients $g_t$ and hessians $h_t$, $t=1,\dots,T$, give the best quadratic approximation to the loss function based on Taylor's theorem, and conveniently allows for deriving a scoring function similar to the impurity score \citep{Breiman.1984} to measure the quality of $f_j$'s tree structure. We utilise the greedy algorithm outlined in \cite{Chen.2016}, which includes imposing a form on $\Omega$ for computational saving strategies.} \add{A natural choice for initialisation is $\hat{\theta}_t^{(0)} \in \arg\min_{\theta}\sum_{t=1}^T  \mathcal{L}(\bm y_t, \theta )$, i.e., a candidate value which, unconditional on the predictors, minimises the empirical loss.}

We choose the number of trees $N$ for each of our boosting sub-models by random five-fold cross-validation, and for parsimony, set the other hyperparameters of our boosting models at values reasonable in our setting (detailed in the Appendix).

\section{Methodology}\label{sec:new_methods}

\subsection{Modelling approach}\label{sec:model:approach}

Our proposed methodology uses generalized $r$-Pareto processes (GrP), described in Section~\ref{sec:grp}, which facilitate the assessment of individual daily functional exceedances of spatial fields. Using an ML algorithm, we develop this framework to incorporate predictors at the same daily resolution as the response. Our spatial model differs from and is an extension of those in univariate EVT such as the GPD, which takes as response exceedances at each grid point. This neither considers spatial dependencies nor provides information on the probabilities that several nearby grid-points are extreme on the same day; this information is crucial for emergency response and adaptation measures.  

While max-stable processes \citep{davison2012} are now commonly used in spatial EVT, 
they are only suitable for modelling grid point-wise maxima fields, which can overlook the co-occurrence of extreme events across multiple grid points on the same day. Furthermore, only modelling maxima fields makes understanding the drivers of heat extremes at fine temporal resolutions such as ours challenging, as models for maxima tend to be only appropriate for coarser resolutions, e.g., monthly or yearly.  

\add{From \eqref{eq:rpareto.convergence}}, statistical inference for $r$-exceedances conditional on a random vector of predictors $\bm X$ is based on the approximation \add{
\begin{align}\label{eq:rPareto:inference}
\operatorname{Pr}(Y \in \mathcal{R} \vert \bm X) 
& = \operatorname{Pr}\left[r\left\{\dfrac{Y-b_n}{r(a_n)}\right\}\geq0 \Bigl\vert \bm X\right] \ \times \operatorname{Pr}\left[ \dfrac{Y-b_n}{r(a_n)} \in \mathcal{R}^\prime \Bigl\vert r\left\{\dfrac{Y-b_n}{r(a_n)}\right\}\geq0, \bm X\right] \nonumber \\ 
& =  \operatorname{Pr}\left[r(Y)\geq r(b_n) \Bigl\vert \bm X\right] \times \operatorname{Pr}\left[ \dfrac{Y-b_n}{r(a_n)} \in \mathcal{R}^\prime \Bigl\vert r(Y)\geq r(b_n), \bm X\right]  \nonumber \\
&\approx {\operatorname{Pr}\left\{r(Y) \geq u_n \vert \bm X  \right\}} \times {\operatorname{Pr}(P^\prime \in \mathcal{R}\vert \bm X)}  ,
\end{align}
where $\mathcal{R} \subset \left\{y \in \mathcal{F}: r(y) \geqslant u_n \right\}$, $u_n=r(b_n)$, $\mathcal{R}^\prime = \left\{y \in \mathcal{F}:  (y - b_n)/r(a_n) \in \mathcal{R} \right\}$ and $P^\prime$ is a GrP with scale function $a_n=A\times r(a_n)$ with location shift $b_n$, i.e., $P^\prime=r(a_n)P + b_n$, where $P$ is the limiting GrP in \eqref{eq:rpareto.convergence} with scale function $A$. The second line of \eqref{eq:rPareto:inference} follows from the linearity of $r$. } 

The schematic overview of our full model and the three sub-models is displayed in Figure \ref{fig:submodels}, and is composed of three sub-models; model one predicts the first term in \eqref{eq:rPareto:inference}, i.e., the probability of an \add{$r$-exceedance}, \rev{which we call the occurrence sub-model}.  Models two and three use the temperature response from the days with an $r$-exceedance to model the last term involving the $r$-Pareto process \add{$P^\prime$}, i.e., the marginal intensity and spatial dependence in the $r$-exceedances of $Y$, \rev{which we call the intensity and spatial dependence sub-models.}

\begin{figure}[t]
\centering
    \includegraphics[width=1\textwidth]{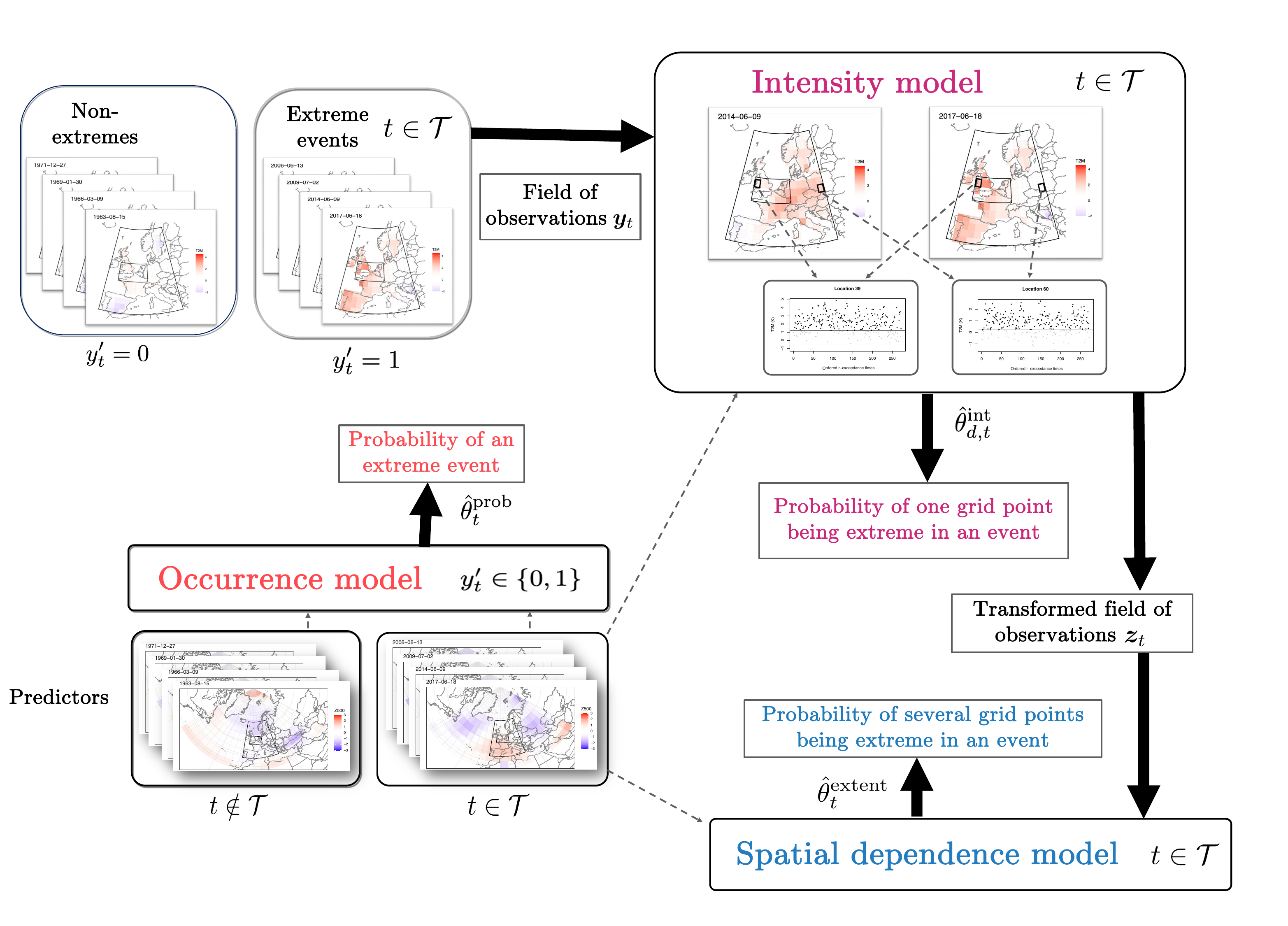}
   \caption{Schematic overview of our sub-models, to complement the description in Section \ref{sec:model:approach}. The occurrence sub-model (in red), takes as input the binary data $y^\prime_t$, for all time points t, and outputs the probability of an $r$-exceedance. The intensity sub-model (in purple), takes as input the the data vector of the T2M response $\bm y_t$, $t\in\mathcal{T}$, and outputs the probability of each grid point exceeding a high threshold at each grid point. The spatial dependence sub-model (in blue), takes as input the transformed field of observations $\bm z_t$, $t\in\mathcal{T}$, and outputs the probability of simultaneous exceedances at several grid points. For the full description of the sub-models, see the text. \add{The parameters $\theta_t^\mathrm{prob}$, $\theta_{d,t}^\mathrm{int}$ and $\theta_t^\mathrm{extent}$ are defined in Section~\ref{sec:model:loss}.} }   
   \label{fig:submodels}
\end{figure}

Our occurrence sub-model first identifies the extreme events, based on the spatial average T2M over all grid points in a \rev{target region $\mathcal{A} \subset \mathcal{S}$}, i.e., we use $r(Y)=\vert\mathcal{A}\vert^{-1} \int_{\mathcal{A}}Y(\mathbf{s}) \text{d}\mathbf{s}$. We choose $\mathcal{A}$ as the region indicated in Figure \ref{fig:t2m_anom} as it encompasses areas with high population density and includes major cities such as London, Amsterdam and Paris. Percentile-based definitions of heat extremes are widely used \citep[e.g.,][]{perkins2015, perkins-kirkpatrick2020}, and we here say that an event is extreme if the spatial temperature average exceeds its $95^\text{th}$ percentile, giving us 272 events. 

More precisely, let $y'_t \in \{0,1\}$ be equal one if there is an $r$-exceedance on day $t$, and zero otherwise, and let $\mathcal{T}$ be the set of days where $y'_t=1$. The occurrence sub-model predicts the occurrence probability of an $r$-exceedance, which uses the binary data $y'_t$ for all days. 

The intensity and spatial dependence sub-models for the last term of \eqref{eq:rPareto:inference} focus specifically on the extreme events, and so only consider the field (in practice, finite-dimensional data vector) of the T2M response for all $t\in \mathcal{T}$. These two sub-models 
consists of an $r$-Pareto model whose marginal distributions and dependence structure can vary with daily predictors. This space-time model can quantify different facets related to the $r$-exceedances' local intensity, e.g., the exceedance probability of the response being large at a grid point, or their spatial dependencies, e.g., the joint probability of the response being simultaneously large at several grid points. To model this last term in \eqref{eq:rPareto:inference}, we use the Brown--Resnick model class detailed in Section~\ref{sec:br}. 

We assume conditional independence given the predictors. This assumes that temporal dependence of the $r$-exceedances is driven solely by the varying predictors, and it withholds the need to decluster our responses in time. We argue that this is reasonable in our case study as our predictors are hypothesised to strongly drive the dynamics of heat extremes, \add{as the SM predictors capture the local land surface effects while the Z500 captures the regional dynamic conditions due to diabatic and adiabatic warming, or advections.}

We use the entire Z500 field (all $242$ grid points) as predictors in all our sub-models. We aggregate the soil moisture predictor spatially over the target region for the occurrence sub-model (see Figure~\ref{fig:composites}B). For the spatial dependence sub-model, we spatially \rev{aggregate} over four approximately equally-sized rectangles (see the four rectangles in Figure \ref{fig:composites}H) over the study region and consider each as a predictor. As the intensity varies in both space and time, we use the latitude, longitude and soil moisture at each grid point as predictors for this sub-model. 

Using these predictors, each sub-model is an ML model that we train \rev{in the distributional boosting framework, described in Section~\ref{sec:boosting}. We do this using loss functions motivated by spatial EVT and the GrP, and we describe these losses next. }

\subsection{Loss functions to model compound extremes}\label{sec:model:loss}
The loss function 
steers the machine learning models that we fit. For the losses we present here, we also derive the closed analytical form of the gradient and the hessian, which facilitates quick computation in the distributional boosting algorithm we use; we compile them in the \texttt{R} code described in the Appendix.  

Let $\bm y_t = (y_{1,t},\dots,y_{D,t})$ be realizations of the daily T2M field $Y$ evaluated at grid points $s_1,\dots,s_D \in \mathcal{S}$ replicated over the days $t=1,\dots,T$, where $T$ is the total number of days in the dataset. Recall that $y'_t \in \{0,1\}$ is the indicator for an $r$-exceedance on day $t$, and that $\mathcal{T}$ contains all the days where $y'_t=1$. 

To model ${\operatorname{Pr}\left\{r(Y) \geq u \vert \bm X  \right\}}$ in \eqref{eq:rPareto:inference}, we consider the response $y'_t$, $t=1,\dots,T$, to predict the probability of occurrence in day $t$: $\mathrm{ilogit}(\theta^\text{prob}_t) \in [0,1]$, where $\mathrm{ilogit}$ is the inverse logit function, i.e., $\mathrm{ilogit}(x) = 1/(1+\exp(-x))$, $x\in\mathbb{R}$. We do this using the log-loss
\begin{align*}
\mathcal{L}_\text{log-loss}(y'_t, \hat{\theta}^\text{prob}_t) =  &y'_t \log\{ \mathrm{ilogit}( \hat{\theta}^\text{prob}_t) \} + (1-y'_t)  \log\{1-\mathrm{ilogit}( \hat{\theta}^\text{prob}_t) \},
\end{align*}
which is standard for modelling binary counts.  

The other two sub-models predict ${\operatorname{Pr}(P \in \mathcal{R}\vert \bm X)}$ in \eqref{eq:rPareto:inference}, where $\mathcal{R} \subset \mathcal{R}\left(u\right)=\left\{y \in \mathcal{F}: r(y) \geqslant u\right\}$. Their fitting procedures proceed sequentially, as is common in copula modelling: we first estimate the marginal parameters and then fit the dependence model after standardising the margins using their predictions from the previous step.

\rev{The approximation in \eqref{eq:rPareto:inference} also involves normalising functions $b_n(\bm s_d; t)$ and \add{$a_n (\bm s_d; t)$}, as detailed in \eqref{eq:rpareto.convergence} of Section~\ref{sec:grp}}. We set the normalising threshold at each grid point to be temporally constant, i.e., $b_n(\bm s_d; t)=b_d$, 
for each $d\in\{1,\dots,D\}$. We follow \cite{deFondeville.2022} and, \rev{due to parameter identifiability constraints,} choose $b_d$ to be the empirical $q'$ quantile at the location $\bm s_d$ from the $r$-exceedances days, with $q'$ chosen such that $r(b_n) = u_n$, where $u_n$ is the $95^\text{th}$ percentile of the empirical distribution of $r(Y)$. We obtain $q'=0.27$ (which still corresponds to extreme values as this is an empirical quantile among the $r$-exceedances), giving $120$ grid points $\times$ $272$ $r$-exceedances $\times (1-0.27) \approx 24'000$ individual excesses across all days and grid points.


The GPD loss \rev{log-}likelihood has garnered recent interest from the EVT modelling community for training ML models \citep[e.g.,][]{velthoen.2021, Koh.2021b, Gnecco.2022}. We predict the scale parameter of the GPD, which we denote 
$\add{a_n (\bm s_d; t)}= \exp(\theta_{d,t}^{\mathrm int})-m_d\xi$, governing the local excesses above $b_d$ at each grid point, by using a variant of this loss, i.e.,
\begin{align*}
    \mathcal{L}_\text{GPD}(y_{d,t}, \hat{\theta}_{d,t}^\text{int}; \xi) &= \mathbbm{1}_{t\in\mathcal{T}} \sum_{d=1}^D \mathbbm{1}_{y_{d,t}>b_d}  \Bigg[ \log\{ \exp(\hat{\theta}_{d,t}^\text{int})  -m_d \xi\} + \dfrac{\xi+1}{\xi} \log\left\{1+\xi \dfrac{y_{d,t}-b_d}{\exp(\hat{\theta}_{d,t}^\text{int})  -m_d \xi}\right\} \Bigg].
\end{align*}
The indicator $\mathbbm{1}_{t\in\mathcal{T}}$ above indicates that only the extreme days contribute to this loss. The parameter $m_d$ is an estimated upper bound that prevents violation of the support constraint of the GPD when $\xi<0$, which is common in environmental data. We set $m_d=\hat{\sigma}_d/\hat{\xi}_d$, where $\hat{\sigma}_d$ and $\hat{\xi}_d$, $d=1,\dots,D$, are the location-wise GPD maximum likelihood estimates of the positive exceedances from $y_{d,t}-\hat{b}_d$, $t\in\mathcal{T}$. An alternative is to set $m_d$ using physical considerations, and we discuss this in Section~\ref{sec:discussion}.

We fix the shape parameter $\xi$ to $-0.3$ due to our five-fold cross-validation study (detailed in the Appendix); this value is also close to the minimum score estimate unconditional on the predictors. A constant shape parameter over $\mathcal{S}$ is not unrealistic here, as we are modelling a type of tail regime of our temperature process governed by large-scale atmospheric circulation and soil moisture. Model diagnostic plots of the fitted model on the test data (displayed later in Figure \ref{fig:qqplot:intensity}) confirm that this assumption is reasonable. Having a spatiotemporally varying shape parameter is possible, and we discuss this more in Section~\ref{sec:grp}, though this approach would entail that we are no longer able to define $r$-exceedances on the scale of the T2M response. 


For the spatial dependence sub-model, we concatenate our predictions and write $\hat{\bm \theta}_{t}^\text{int} = (\hat{\theta}_{1,t}^\text{int},\dots, \hat{\theta}_{D,t}^\text{int} )$ and $\hat{\bm b} = (\hat{b}_1,\dots,\hat{b}_D)$. We transform our observations to $\bm{z}_t = \{1+ (\bm{y}_t-\hat{\bm b})\xi/\exp(\hat{\bm \theta}_t^\text{int})\}_+^{1/\xi}=(z_{1,t},\dots,z_{D,t})$, $t=1,\dots,T$, and use these to predict a measure of the spatial extent $\theta^\text{extent}_t\in \mathbb{R}$ from the \add{semivariogram that models the dependence of} the Brown-Resnick model from \eqref{eq:covariance} in Section~\ref{sec:br} using the loss
\begin{align}\label{eq:rparetoloss}
\mathcal{L}_\text{GrP}(\bm{z}_t, \hat{\theta}^\text{extent}_t) &= \mathbbm{1}_{t\in\mathcal{T}} \sum_{d=1}^D \left\{ w_d(z_{d,t})^2 \dfrac{\partial \lambda_\theta(\bm{z}_t) }{\partial z_{d,t}} \Bigl\vert_{\theta=\hat{\theta}^\text{extent}_t} \right\} + 2 \dfrac{\partial}{\partial z_{d,t}}\left\{ w_d(z_{d,t})^2\dfrac{\partial \lambda_\theta(\bm{z}_t) }{\partial z_{d,t}}\Bigl\vert_{\theta=\hat{\theta}^\text{extent}_t} \right\},
\end{align}
where $w_d: \mathbb{R}_+ \rightarrow \mathbb{R}_+$ is the positive weight function derived in \cite{HYVARINEN.2007}, and $\lambda_\theta$ is the intensity function \add{\cite{wadsworth14} 
\begin{align*}
\lambda_\theta(x) = &\frac{|\det \boldsymbol\Sigma_\theta^*|^{-1/2} (1^\top_D \rho)^{-1/2}}{(2\pi)^{(D-1)/2} x_1 \cdots x_D} 
\exp \left( - \frac{1}{2} \left[ \log x^\top \boldsymbol\Gamma \log x + \log x^\top \left\{ \frac{2\rho}{1^\top_D \rho} + (\boldsymbol\Sigma_\theta^*)^{-1} \boldsymbol{\sigma} - \frac{\rho \rho^\top \boldsymbol{\sigma}}{1^\top_D \rho} \right\} \right] \right) \\
&\times \exp \left[ - \frac{1}{2} \left\{ \frac{1}{4} \boldsymbol{\sigma}^\top (\Sigma_\theta^*)^{-1} \boldsymbol{\sigma} - \frac{1}{4} \frac{\boldsymbol{\sigma}^\top \rho \rho^\top \boldsymbol{\sigma}}{1^\top_D \rho} + \frac{\boldsymbol{\sigma}^\top \rho}{1^\top_D \rho} - \frac{1}{1^\top_D \rho} \right\} \right], \quad x \in \mathbb{R}_+^{D},
\end{align*} 
where $1_D$ is a $D$-dimensional vector with unit components, $\boldsymbol \Sigma^*_\theta$ is the $D$-dimensional covariance matrix of the intrinsically stationary Gaussian process with semivariogram $\gamma$ given in \eqref{eq:covariance}, $\rho = (\boldsymbol\Sigma^*_\theta)^{-1} 1_D$, $\boldsymbol\Gamma = (\boldsymbol\Sigma^*_\theta)^{-1} - {\rho \rho^\top}/{1^\top_D \rho}$ and $\boldsymbol{\sigma} = \mathrm{diag}(\boldsymbol\Sigma^*_\theta)$. Computing the gradients and hessians of the intensity function with respect to $\theta$ (needed in \eqref{eq:objective}) and $x$ (needed in \eqref{eq:rparetoloss}) is straightforward, though tedious; we provide their explicit forms in the \texttt{R} code in the Appendix.}

Equation \eqref{eq:rparetoloss} is based on the gradient scoring rule \cite{Hyvarinenscore05} derived for the Brown-Resnick model in \cite{deFondeville.2022} to circumvent computation of the GrP likelihood which is quickly intractable for moderate $D$. Here we adapt the gradient score to incorporate temporal non-stationarity via predictors.

We here fix $\alpha$ and ${\theta}^\text{scale}$ from Section~\ref{sec:br} to be temporally constant at 1.27 and -0.07, which are their corresponding minimum gradient score estimates from the model fitted without the predictors. One could also predict $\theta_t^\text{scale}$ in our algorithm by replacing $\hat{\theta}_t^\text{extent}$ by $\hat{\theta}_t^\text{scale}$ in \eqref{eq:rparetoloss}, and apply our fitting procedure consecutively between the two parameters. However, we choose a more parsimonious approach in our data-scarce setting and only let ${\theta}_t^\text{extent}$ vary with predictors.  
 

The numerical complexity of \eqref{eq:rparetoloss} is that of matrix inversion, i.e., $\mathcal{O}(D^3)$. To facilitate quicker computation, 
we construct numerically convenient representations of Gaussian processes using the Vecchia approximation \citep{Vecchia.1988} with $k$ conditioning variables. This sparsifies the Cholesky factor of the precision matrix to reduce the computational cost to $\mathcal{O}(Dk^3)$ when $k\ll D$. Here we use the ordering outlined in \cite{Guinness.2018}, which has \add{been shown in a two-dimensional spatial setting} to improve on the lexicographic one originally proposed, and our numerical results suggest that taking $k\approx 20$ is reasonable. \add{This ordering is called an approximate maximum minimum distance approach, where a point in the center of the space is chosen first, and then each successive point is chosen to maximize the minimum distance
to previously selected points.} Detailed in the Appendix, we check by simulation that the parameters are identifiable and can be predicted adequately in a setting similar to ours. Such verification is especially important for the spatial dependence sub-model due to its complexity, and since detecting non-stationarity in the extremal spatial dependence structure is challenging with limited data. 

\section{Results}\label{sec:results}

\subsection{Model simulation and evaluation on the test set}

A key feature of our ML model grounded in EVT is that it is 
generative. It can simulate a range of extreme weather scenarios by varying predictor conditions and also to explore theoretically-driven alternative scenarios under fixed predictor conditions, thereby extending the model's utility beyond prediction.

\begin{figure}[t]
    \centering
\includegraphics[width=.7\textwidth]{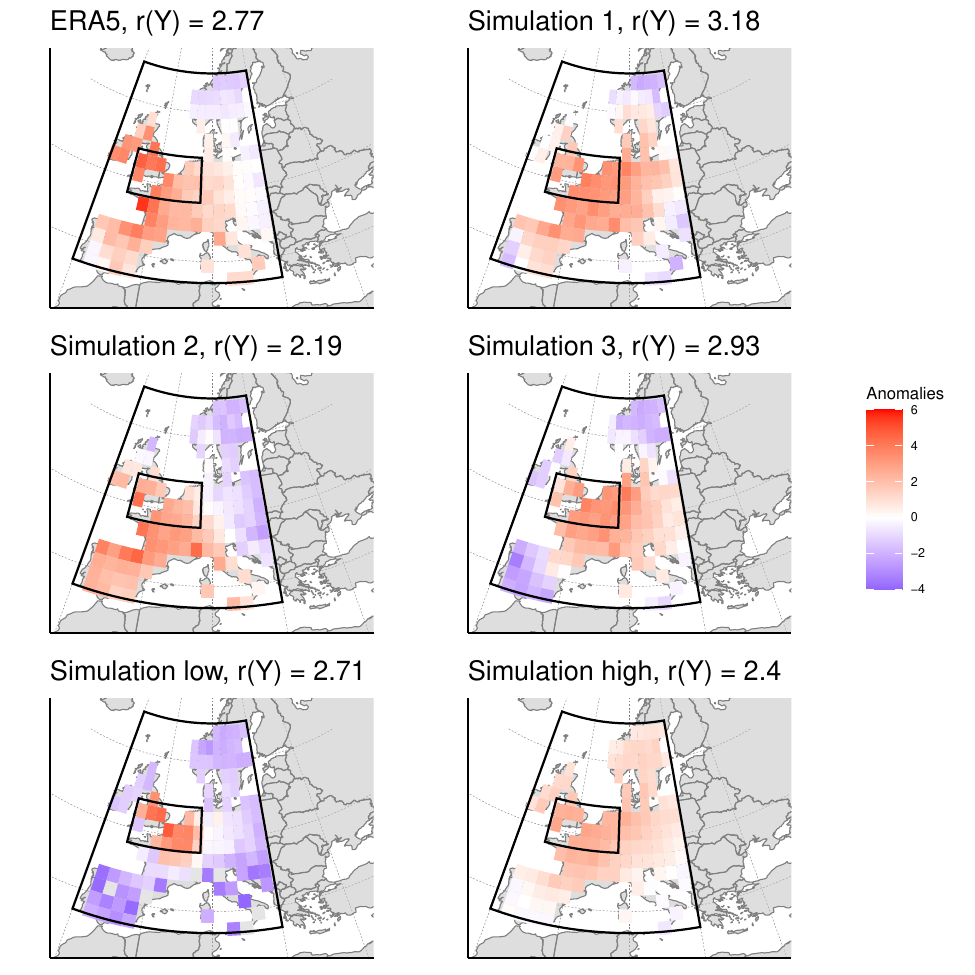}
  \caption{ERA5 realisation of the surface temperature anomaly (top left panel) and model simulations (top right and middle panels) on 18 July 2022 given an $r$-exceedance. The subtitle in each panel indicates the spatial average of T2M over all grid points in the target region. The bottom panel show simulations from a GrP with all the same model-predicted parameters except for $\theta_t^\text{extent}$, which we artificially set to a low of 0.05 and a high of 4; our model instead predicted $\hat{\theta}_t^\text{extent}=1.81$. A grey fill at a grid point in the study region indicates an anomaly below -4K. }
  \label{fig:modelsim}
\end{figure}

To showcase this, 
we focus on the European heat wave 2022 (Figure \ref{fig:intro_2022}) from the test set. On 18 July 2022, an $r$-exceedance event occurred. We use the predictors for that day to simulate alternative heat extremes based on predictions from our intensity and spatial dependence sub-models. Figure \ref{fig:modelsim} (top four panels) presents three such alternative $r$-exceedance scenarios and the ERA5 realisation of the event. 
Importantly, all simulated scenarios from the fitted model exhibit an intensity and a spatial extent that is visually similar to both the observed July 18 event (top left panel of Figure \ref{fig:modelsim})  and other regional heat extremes (not shown), while the bottom two panels that show simulations from a GrP with \add{artificially} low and high dependence parameters do not; an analysis summarising the satisfactory fit of the model is given in the Appendix (Figure \ref{fig:extremogram}).

\begin{figure}[t]
    \centering
\includegraphics[width=.35\textwidth]{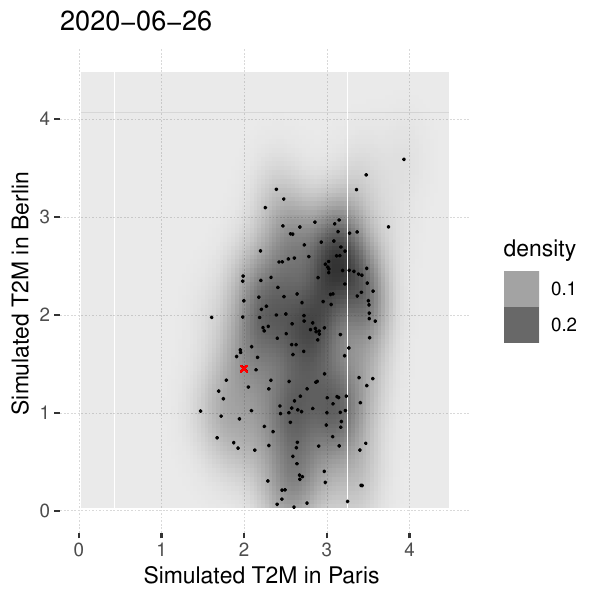}
\includegraphics[width=.35\textwidth]{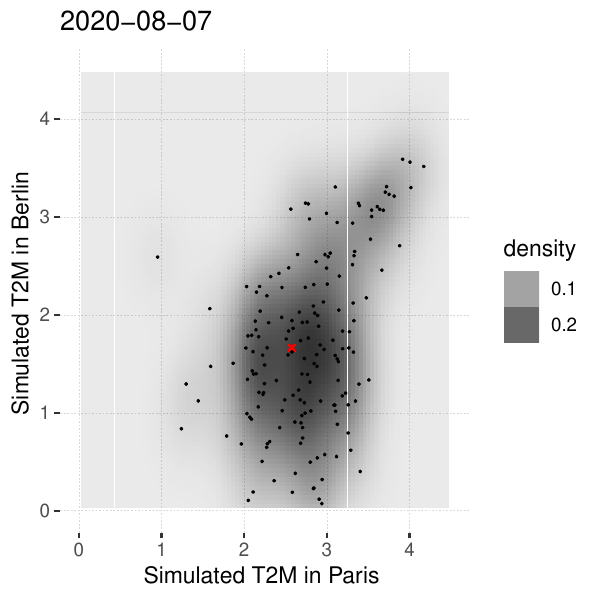}
  \caption{Two-dimensional density plots of T2M values at two grid points (containing Paris in the x-axes and Berlin in the y-axes) from 200 model simulations (black dots) on two heat extreme days (represented by the two panels) in the test set, conditional on positive T2M anomalies and an $r$-exceedance. The red crosses represent the observed T2M from the ERA5 reanalysis.}
  \label{fig:modelsim_dep}
\end{figure}


Figure \ref{fig:modelsim_dep} showcases 200 model simulations for \rev{two} specific heat extreme days affecting both Berlin and Paris. The observed T2M at both grid points on both days are within the range of the heat events generated by the model. Regarding process understanding, these simulations suggest that even more severe events with jointly high T2M values at multiple grid points are plausible. 
This illustration highlights the strength of our theoretically-justified approach to extrapolate beyond the range of observed data with plausible \rev{intensities and} spatial extremal dependencies.

To evaluate our models' performance, we compare our predictions to ERA5 on the 
test set (2018--2022) (Figure \ref{fig:roc}). The occurrence sub-model obtains an impressive area under the curve score (AUC) of 0.899, and its Receiver Operating Curve (ROC) is displayed in Figure \ref{fig:roc}. Notably, the Z500 predictors substantially improve the predictive performance of the occurrence sub-model.  For the intensity sub-model, Figure \ref{fig:qqplot:intensity} highlights a good overall fit to the local empirical tail distribution for an $r$-exceedance at four selected grid points containing Berlin, London, Barcelona and Paris. 
The good predictive skill when modelling spatial dependence in extremes is discussed in the Appendix. 

\begin{figure}[t]
    \centering
\includegraphics[width=.45\textwidth]{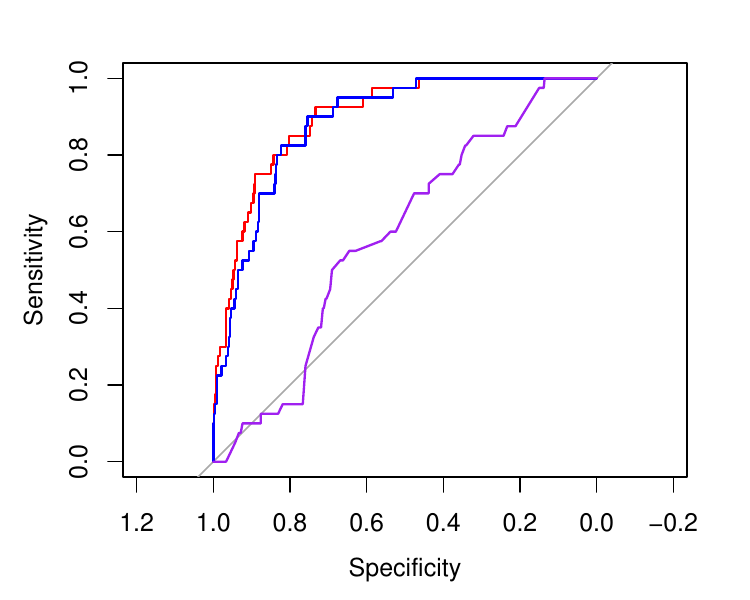}
  \caption{
  ROC for the occurrence sub-models on the test set (2018--2022). Displayed are the curves corresponding to the model without the SM predictor (blue), without the Z500 predictor (purple), and the full model (red). The curve closer to the top left corner (and away from the diagonal) is better. The Area under the Curve (AUC) scores are 0.589, 0.885 and 0.899, respectively. 
  }
  \label{fig:roc}
\end{figure}

\begin{figure}[t]
    \centering
        \begin{subfigure}{0.3\textwidth}
\includegraphics[width=.95\textwidth]{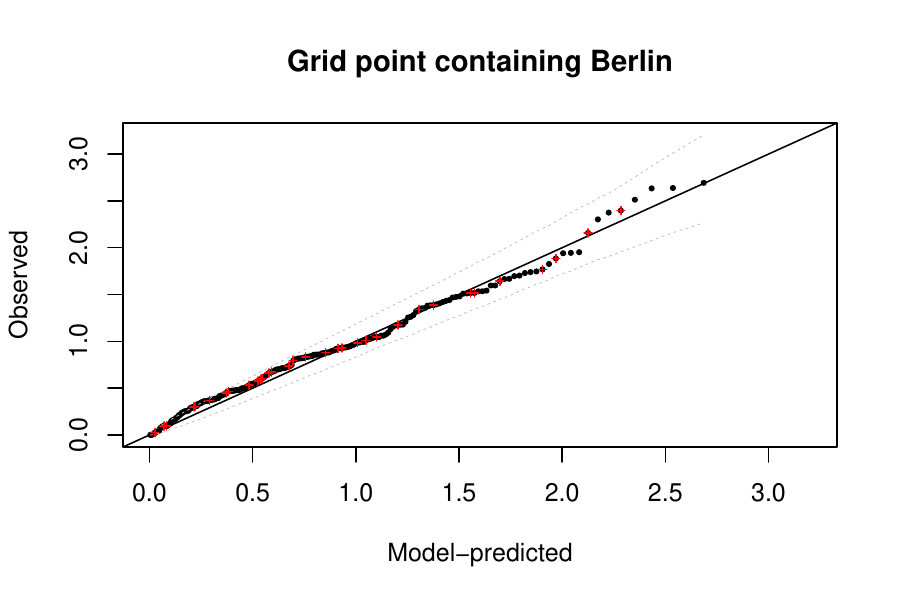}
\end{subfigure}
\begin{subfigure}{0.3\textwidth}
\includegraphics[width=.95\textwidth]{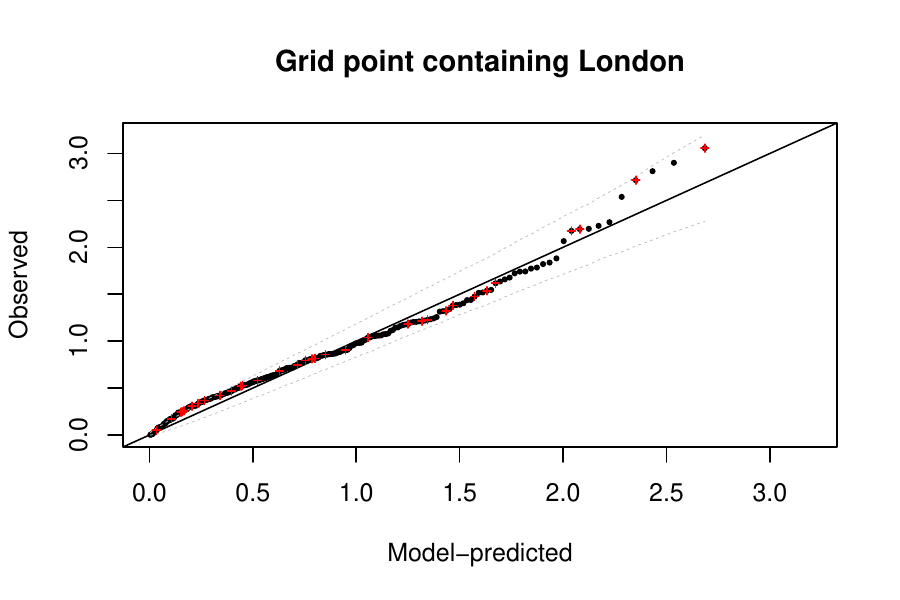} 
\end{subfigure}\\
 \begin{subfigure}{0.3\textwidth}
\includegraphics[width=.95\textwidth]{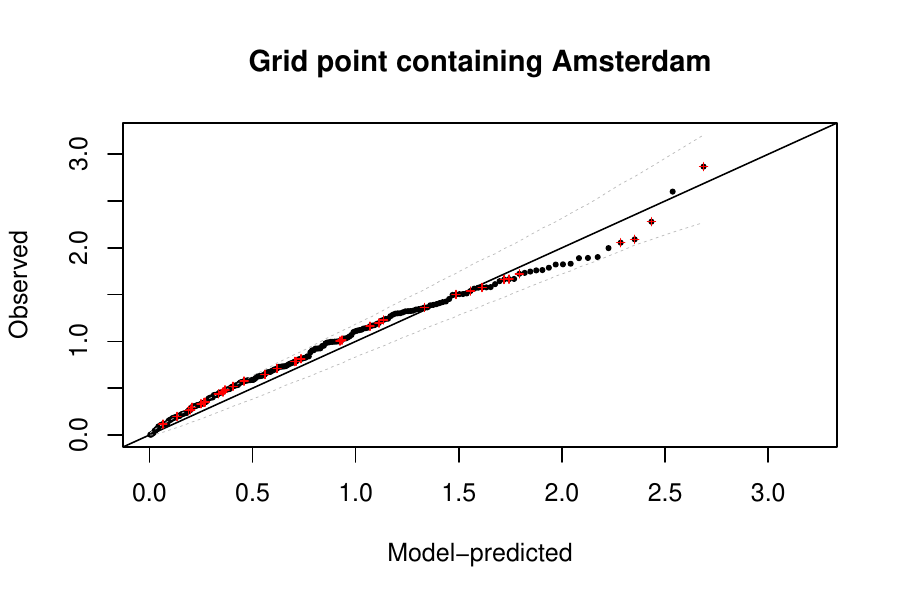}
\end{subfigure}
\begin{subfigure}{0.3\textwidth}
\includegraphics[width=.95\textwidth]{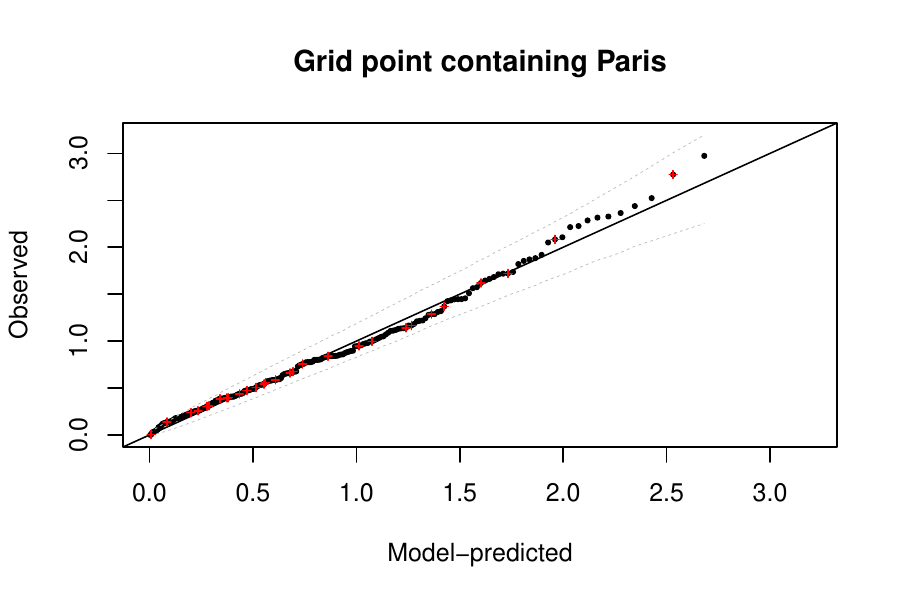} 
\end{subfigure}
  \caption{Quantile-Quantile plots comparing the intensity model's scaled tail distributional prediction above $b_d$ given an $r$-exceedance (x-axes), 
  with that from ERA5 (y-axes), 
  for four chosen grid points $d$ and on both the training (black dots) and test sets (red crosses). The dashed lines correspond to $95\%$ pointwise parametric bootstrap confidence intervals. } 
  \label{fig:qqplot:intensity}
\end{figure}

\subsection{Importance of Z500 large-scale weather patterns}

We use the predictions from our model to identify key features of the large-scale weather patterns associated with the formation of European heat extremes. 


We analyze composite maps of the Z500 predictor field conditioned on when each sub-model predicted high values, i.e., the top 10\% of predicted occurrence probabilities, the intensities, and the spatial dependence of the heat extremes in the training set. Additionally, \add{SHapley Additive exPlanation (SHAP) values, and in particular, estimates of SHAP values for tree-based models \cite{Lundberg.2019} that we briefly describe in the Appendix, give a measure of the contribution of each Z500 predictor to the final outcome of each sub-model.} 


The Z500 values associated with the top 10\% of values in all three sub-models (Figure~\ref{fig:composites}A,D,G) show anticyclonic anomalies (positive anomalies exceeding 1.5 standard deviations) over the target region. The positive anomalies are part of a large-scale wave pattern, 
which includes an upstream trough over the Atlantic Ocean (negative anomaly) and a weaker downstream trough over eastern Europe (negative anomaly). Anticyclonic circulation anomalies favour subsidence, leading to adiabatic heating and cloud-free skies associated with an absence of precipitation. 
In addition, the upstream trough can contribute to positive horizontal air advection at the upstream flank of the anticyclone \citep[e.g., ][]{kautz2022,Tuel.2023.EGUSphere}. 

Comparing the composites for occurrence probability (Figure~\ref{fig:composites}A), intensity (Figure~\ref{fig:composites}D), and spatial dependence (Figure~\ref{fig:composites}G), we note differences in the relative position, the intensity, and the shape of the anticyclonic anomaly and of the troughs across the composites. 
For the occurrence model (Figure~\ref{fig:composites}A), the trough (negative anomaly) over the Mediterranean to the south of the anticyclonic anomaly indicates the presence of a dipole blocking pattern, favoring the stagnation of the large-scale flow \citep{rex1950}. The SHAP metric in Figure~\ref{fig:composites}C highlights the centre of the anticyclonic anomaly and the region upstream over the North Atlantic as important - emphasizing the importance of both the local and the upstream circulation patterns for heat extreme occurrence.
For the intensity model (Figure~\ref{fig:composites}D,F), the anticyclonic anomaly is more elongated in the meridional direction and is flanked by upstream and downstream troughs, resembling an omega blocking pattern. The SHAP analysis does not highlight a specific area within the block, but rather suggests that the heat intensity is influenced by the entire trough-ridge pattern across the Euro-Atlantic region.
For the spatial dependence model (Figure~\ref{fig:composites}G), the Z500 composite shows a zonally broader anticyclonic anomaly, and one that is less intense compared to those in the other composites. The SHAP analysis (Figure~\ref{fig:composites}I) draws attention to the anticyclone's flanks, underscoring the zonal extent of the anticyclonic anomaly, rather than its core intensity, as important in driving spatially extensive heat extremes.

\begin{figure*}[t]
    \centering
\includegraphics[width=.99\textwidth]{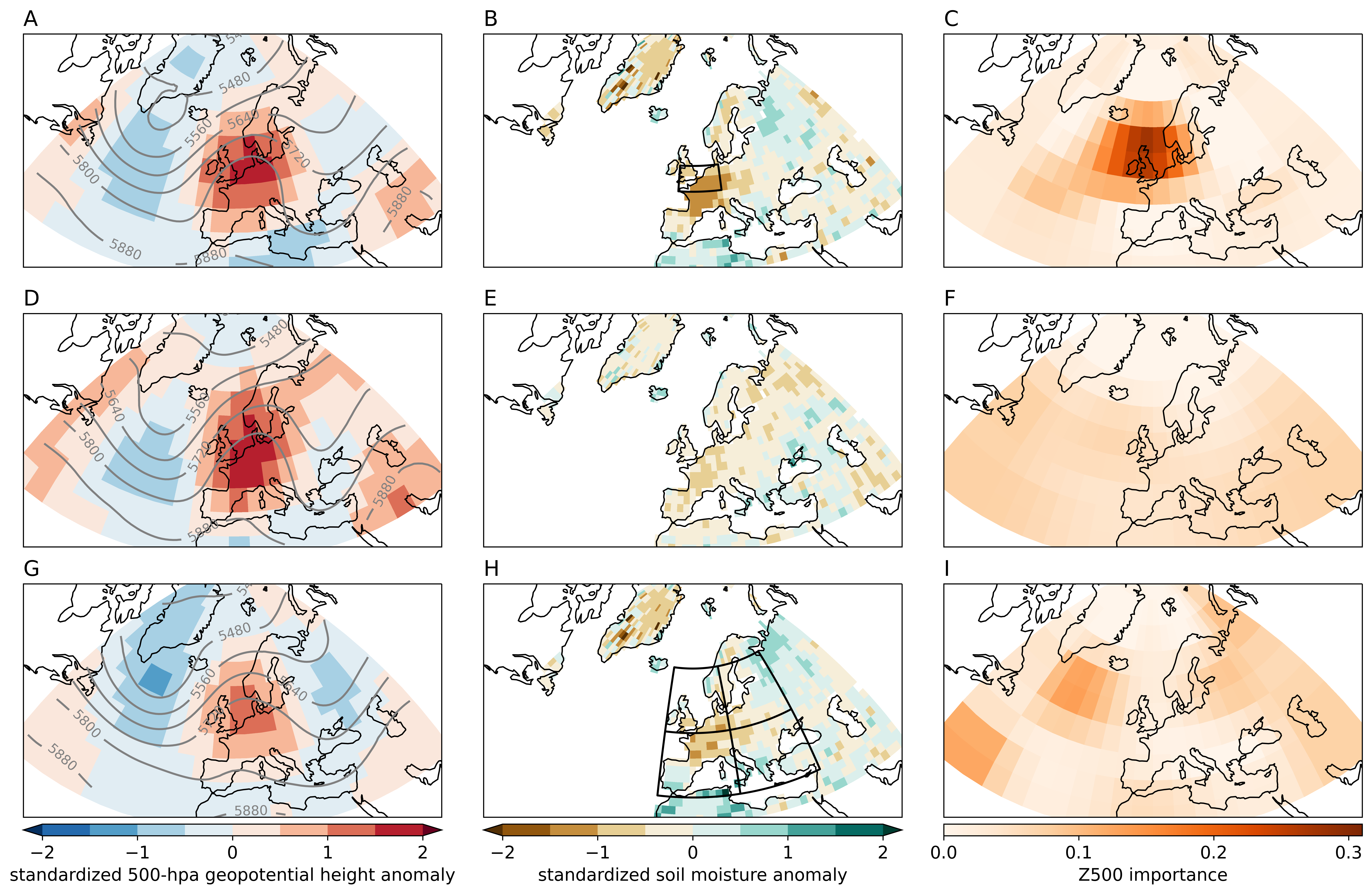}
  \caption{Composite maps of Z500 (color shading), Z500$^\prime$ contours (grey lines)(A,D,G), and maps of SM (B,E,H) 
  for the 10\% heat extremes with highest predicted occurrence probabilities (first row), intensity (second row), and spatial dependence (third row). Variable importance plots (with kernel density spatial smoothing) of the top Z500 predictors with the region averaged SHAP metric for the (C) occurrence, (F) intensity, and (I) spatial dependence sub-models. The rectangles represent the regions used to aggregate the SM predictor for use in the occurrence (B) and spatial dependence (H) sub-models. }  
  \label{fig:composites}
\end{figure*}

\subsection{Importance of soil moisture}

Soil moisture anomalies emerge as a salient predictor across our models, albeit with varying impacts on the performance. SM align well with the Z500 composites, particularly in their spatial distribution: SM is depleted beneath the anticyclone (Figure~\ref{fig:composites}B,E,H).

In the occurrence sub-model, SM averaged over the target region ranks sixth in importance among 243 predictors based on SHAP values. Beneath the anticyclone, SM is notably depleted (more than -1 standard deviation away from 1991--2020 climate, Figure \ref{fig:composites}B), implying that (pre-conditioning) soil dryness (in our case, over the last 15 days) is favourable for heat extremes, in accordance with the literature \citep[e.g.,][]{fischer2007, miralles2014}. Model performance, as evidenced by the ROC, improves slightly with the SM predictor in the test set (Figure \ref{fig:roc}), though the role of the Z500 predictors still dominates.



For the intensity sub-model, the local SM at a specific grid point appears as the top three most important predictors (along with latitude and longitude). SM is depleted in the target region (Figure \ref{fig:composites}E), albeit less compared to the occurrence sub-model's composite map. Unlike before, however, the inclusion of the SM predictor does not significantly improve test error, as measured by the Generalized Pareto Distribution (GPD) loss likelihood (not shown). This suggests that extreme heat intensities can be simulated by our model just as accurately without SM information.

In the spatial dependence sub-model, the SM is depleted below the anticyclone, but not in its surrounding areas, as shown in Figure \ref{fig:composites}H. The predictor corresponding to the average SM over the United Kingdom (top left region of our study region) appears in the top 10 spatial dependence sub-model, though as before, the predictive improvement of having the SM related predictors is small.

\section{Discussion and conclusion}\label{sec:discussion}

A key ingredient of machine learning models is the loss function used for training, and choices for these functions in climate sciences have largely been restricted to those that emphasise good prediction of the distributional bulk rather than the tails. For example, the root mean squared error, the default for modelling real-valued responses in many applications, implicitly presupposes that the target of interest is the conditional mean given the predictors, which may be inappropriate if the focus is on extreme values. 
We propose a new approach for modelling spatially compounding weather extremes that focuses specifically on the distributional tails of the data. Our approach uniquely employs loss functions motivated by spatial EVT to assess daily extreme weather events. 

An important feature of our approach is the ability to extrapolate beyond the range of the observational record, not just for the magnitude but also regarding the spatial extent of extreme events. We showcase this in our case study by using the model to generate alternative extreme heat events in our test set. Our approach provides an alternative to physical model-based techniques or other ML models. This sets a solid foundation for future work with spatially compounding extremes where the impacts crucially depend on the spatial extent of the events, such as for extreme precipitation that leads to floods, or for dry and hot conditions that result in wildfires.


For model validation, we use daily gridded data from ECMWF's ERA5 reanalysis, recognizing that the limited duration of observational records presents challenges for extreme-value analysis and evaluating tail properties in the data distribution. Future work may therefore consider leveraging long-term climate simulations \citep[e.g., as done in][]{ragone.2018, zeder2023}, to obtain more robust model validation. 

Unlike traditional approaches that employ summary metrics of 500-hpa geopotential height spatial patterns, such as EOF-derived weather regimes \citep{grams2017, rouges2023} or selected regional averages, 
our approach leverages predictors \add{at high spatiotemporal resolution} to model the functional exceedances of spatial fields. Applied here to the study of heat extremes in Western Europe, this allows for a more granular understanding of the large-scale processes associated with their occurrence, intensity, and spatial dependencies. 

Our predictors capture persistence and pre-conditions, as they are averaged over the previous 5 days (Z500) or 15 days (SM). We choose these predictors based on physical mechanisms known to contribute to heat extremes. Our model elucidates the large-scale weather patterns conducive to heat extremes in Western Europe. Heat extremes in Europe are closely linked to 500-hpa geopotential height anomalies of the atmospheric circulation with intense (blocking) anticyclones and upstream and downstream troughs. (Blocking) anticyclones provide a condition favourable for the build-up of high temperature over the study region \citep[e.g.,][]{rothlisberger2023, barriopedro2023}. Anticyclones extending farther north are found to be associated with intense heat, while broader anticyclonic anomalies are associated with spatially extended heat extremes. 
We find that upstream and downstream troughs are also important in modulating the intensity and spatial extent of heat extremes, confirming the findings of \citep{steinfeld2020, neal2022}. 

Recent record-breaking heat events have raised questions on the reliability of estimates arising from classical EVT-based analysis. An approach that incorporates upper-tail end information based on physical considerations \citep[e.g., ][]{noyelle:hal.2024} could be implemented as an extension to our model.  

\rev{There is general agreement that \add{preceding} soil moisture deficits can exert a positive feedback on heat waves through modulation of the energy fluxes at the surface in areas where soil moisture is in a transitional regime \citep[e.g.,][]{seneviratne2010,Martius.2021}. \cite{Whan.2015} find an effect of soil moisture on heat wave intensity. The question of soil moisture effects on heat wave extent has to our knowledge not yet been studied in detail. We here find that} pronounced dry soils beneath the (blocking) anticyclone prior to the heat extremes are important for the occurrence of heat extremes,  
though its importance as a precursor for the intensity and spatial dependence of heat extremes is less clear. Soil moisture depletion might also be more relevant during heat extremes \citep{Pappert.2024}. 

\add{The spatial scales of the weather systems that are typically associated with dry soil conditions \citep{Pappert.2024} can justify our assumption of a constant soil moisture content within a grid point. Land-atmosphere interactions however, are highly complex and choosing soil moisture as the only predictor to represent these interactions is a strong simplification, as we neglect the effects of other factors like soil characteristics, surface characteristics, vegetation characteristcs and irrigation. Future work could incorporate other such spatial complexities into a model, especially if an application requires a more refined representation of land-atmosphere interactions or the representation of effects like Urban Heat Islands.}





\add{We here assumed conditional independence of the response given the predictors, which is a common assumption in the ML literature. Future work could investigate statistical testing procedures that could validate this assumption in our framework. We posit that a violation of this assumption will not affect the unbiasedness of our predictions but rather affect their variances, as it is typical to lose statistical efficiency of estimators using a misspecified loss or likelihood.}

\add{The GPD was used as a model in our study to estimate the marginal parameters in the GrP process, even with the additional conditioning event $\{r(Y)\geq u\}$ which restricted us to consider as data response the extreme $r$-exceedance days. One could question the validity of the GPD as the limiting model, especially at subasymoptic levels. In these scenarios, we nevertheless view the GPD as a flexible two-parameter model that is able to capture marginal tail behaviour.  Other methods, such as those based on probability weighted moments or on other distributions, could also be considered in future work. }

\add{Relaxing the assumption of a constant shape parameter $\xi$ is possible, possibly leveraging the model in \citep{Dombry-Ribatet.2015}. However, to avoid losing the physical interpretation of the risk in terms of the original data, current theoretical bottlenecks still restrict us to using generalized $r$-Pareto processes with a constant $\xi$.}

\add{One could utilise the predictors in machine learning models for the thresholds $b_n$ and $u_n$, perhaps using methods like quantile random forest \citep{Meinshausen.2006}. Although out of scope of our work, this approach remains an interesting research avenue. }

\add{Recent generative models \citep[e.g., those reviewed in][]{shi2025.deep} could generate Z500 fields given certain boundary conditions (including SM). With our modelling approach, one could in principle predict $\operatorname{Pr}(Y\in \mathcal{R} )$, the unconditional version of \eqref{eq:rPareto:inference} using simulated fields from such models. }

This work motivates new avenues of research using ML tools to fit new theoretically-motivated models for extreme values. This includes but is not limited to, varying types of extremes, regions of interest, predictor sets, and even the types of machine learning models deployed.  Another related avenue is to use our framework to improve on numerical weather prediction or sub-seasonal to seasonal prediction models, serving as an advanced postprocessing tool.




\enlargethispage{20pt}

\ack{The authors gratefully acknowledge support from Johanna Ziegel, the Oeschger Centre for Climate Change Research and the Wyss Academy for Nature, University of Bern. \add{Further thanks goes to the two anonymous referees and Raphael de Fondeville for their constructive comments that improved the quality of this paper. OM acknowledges support from the Swiss National Science Foundation Grant Number 200020\_207384. }}


\vskip2pc

\bibliographystyle{RS} 

\bibliography{sample} 

\newpage

\section{Appendix}

\subsection{Simulation study}
We simulate from a Brown-Resnick model with the same number of $r$-exceedance times and grid points as in our case study, with $\alpha=1$ and ${\theta}^\text{extent}_t$ 
depending only a few Z500 predictors from the training set, i.e., at selected grid points, in a non-linear way. We do not detail the precise form of this dependence here but rather detail it in the \texttt{R} code replicating this simulation. We still use the whole field of Z500 predictors when training the model to examine if our learning procedure can discriminate the signal from the noise.


We predict all the parameters for 100 independent simulated replicates of this model. 
Figure \ref{fig:sim} suggests that the trend is already well detected with $272$ $r$-exceedances, even when the initial starting values were far from the truth (left panel). Given sufficient boosting iterations (190 in our example), the predicted joint probabilities of exceedance are generally unbiased with $42\%$ of the truth falling within the interquantile ranges of the boxplots in the right panel. The empirical variance tends to be larger for the ones associated with higher joint probabilities; this is consistent with previous studies that found that the parameters associated with higher spatial extended events tend to be harder to estimate \citep[e.g.,][]{sang2014}. 

\begin{figure}[ht]
\centering
    \includegraphics[width=.45\linewidth]{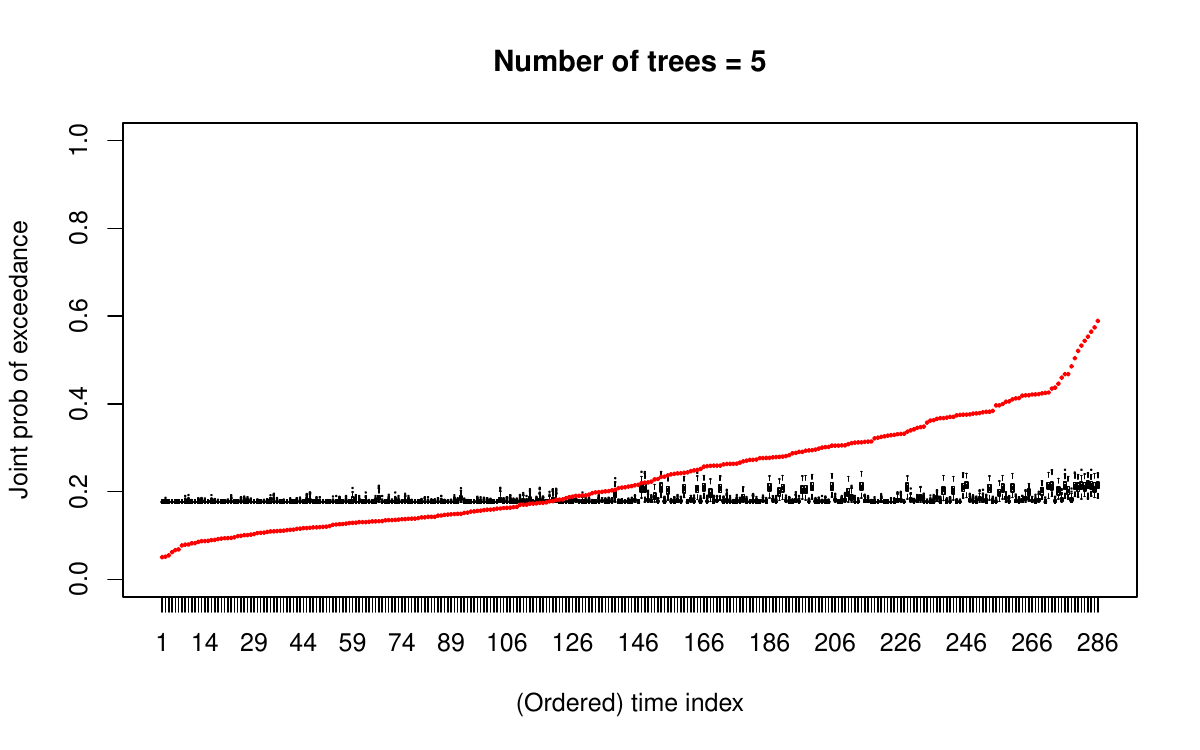}
    \includegraphics[width=.45\linewidth]{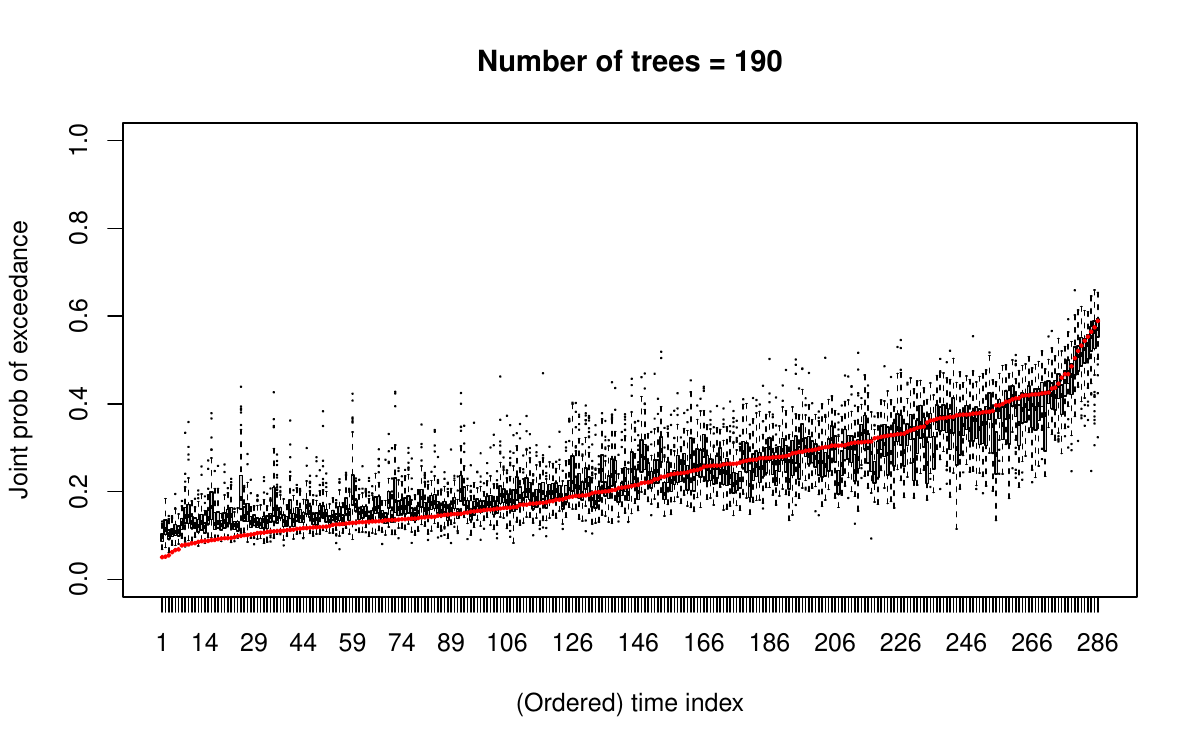}
  \caption{Boxplots for $\hat{\pi}_t(\bm s_1, \bm s_2)$ in our simulation study after 5 (left) and 190 (right) boosting iterations, where $\bm s_1$ is the location of London and $\bm s_2$ of Barcelona. The red dots represent the simulated truth.}   
  \label{fig:sim}
\end{figure}

\subsection{Boosting hyperparameters and cross-validation}

We choose the maximum depth of the trees used in all our sub-models to be between $5$ to $8$, and the stepsize at each boosting iteration as $0.05$ in all cases. Our boosting algorithm is based on the approximate algorithm with weighted quantile sketch in \texttt{xgboost}; for more details see \citep[][Appendix A]{Chen.2016}. Figure \ref{fig:CV_M} shows the five-fold cross validation results from the spatial dependence and intensity sub-models. For the intensity sub-model, $\xi=-0.3$ was chosen because the cross-validation curve corresponding to that setting gave the lowest score. 

\begin{figure}[ht]
    \centering
\includegraphics[width=.48\textwidth]{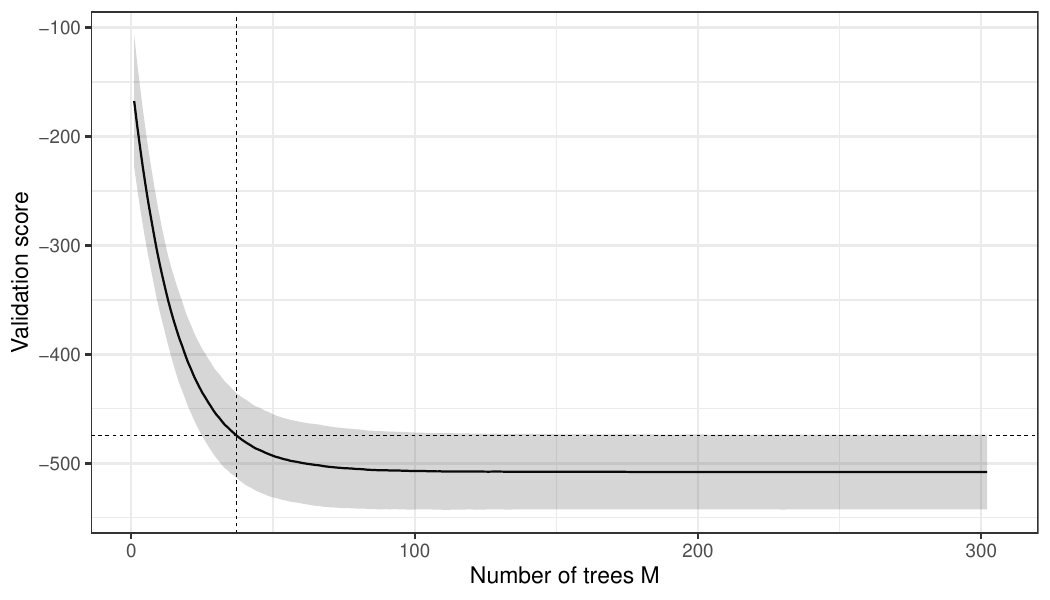}
\includegraphics[width=.48\textwidth]{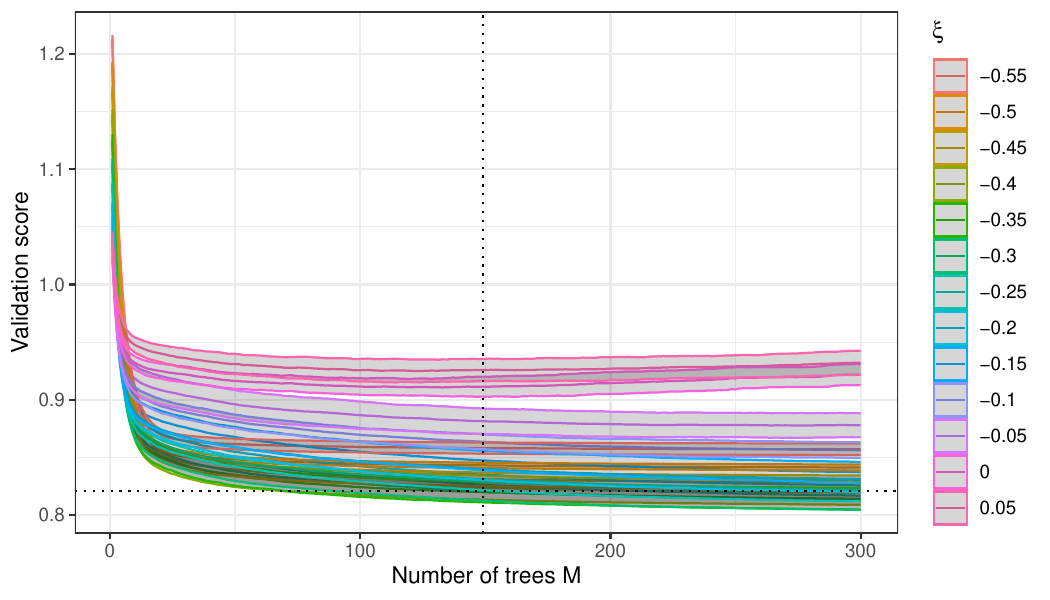}
  \caption{The average rescaled evaluation score across a random five-fold cross-validation scheme for the (left) spatial dependence model using $\mathcal{L}_\text{GrP}$ to predict $\hat{\theta}_t^\mathrm{extent}$, and (right) the intensity model that uses $\mathcal{L}_\text{GPD}$ to predict $\hat{\theta}_{d,t}^\mathrm{int}$. The shaded regions shows the pointwise one-standard-error bound. The dotted lines pinpoints the number of trees chosen by the one-standard-error rule. }
  \label{fig:CV_M}
\end{figure}

\subsection{Tree-based SHAP}\label{sec:shap}

\rev{
The Shapley value \cite{Shapley.1953} was introduced in game theory to fairly distribute a collective payout among players. In machine learning contexts, each predictor is a player and the payout the prediction. Shapley values quantify the contribution of predictor $X_j$ in a model via its marginal contribution to model prediction averaged over all possible models with different combinations of predictors, i.e., 
\begin{equation}\label{eq:shapley}
    \sum_{S\subseteq N \setminus \{j\} } \frac{k!(p-k-1)!}{p!}\left\{f(S \cup \{j\}) - f(S) \right\},
\end{equation}
where $p$ is the total number of predictors, $N\setminus\{j\}$ is a set of all possible combinations of predictors excluding $X_j$, $S$ is a feature set in $N\setminus\{j\}$, $f(S)$ is the model prediction with predictors in $S$, and $f(S \cup \{j\})$ is the model prediction with predictors in $S$ and predictor $X_j$. Though \eqref{eq:shapley} is theoretical optimal, it is difficult to compute and approximations have been proposed, e.g., SHapley Additive exPlanations \citep[SHAP; ][]{Lundberg.SHAP.2017}. We use the fast SHAP value estimation method proposed in \citep{Lundberg.2019}, which is tailored for ensemble tree-based predictions like those arising from gradient tree boosting or random forest models.  }

\subsection{Added value of the spatial dependence sub-model}

\rev{Figure \ref{fig:extremogram} shows that the model provides much better fit to the empirical estimates of $\eqref{eq:condprob}$ than simulations low and high (from Figure \ref{fig:modelsim}) in the test set. }

\begin{figure}[ht]
    \centering
 \begin{subfigure}{0.5\textwidth}
\includegraphics[width=.9\textwidth]{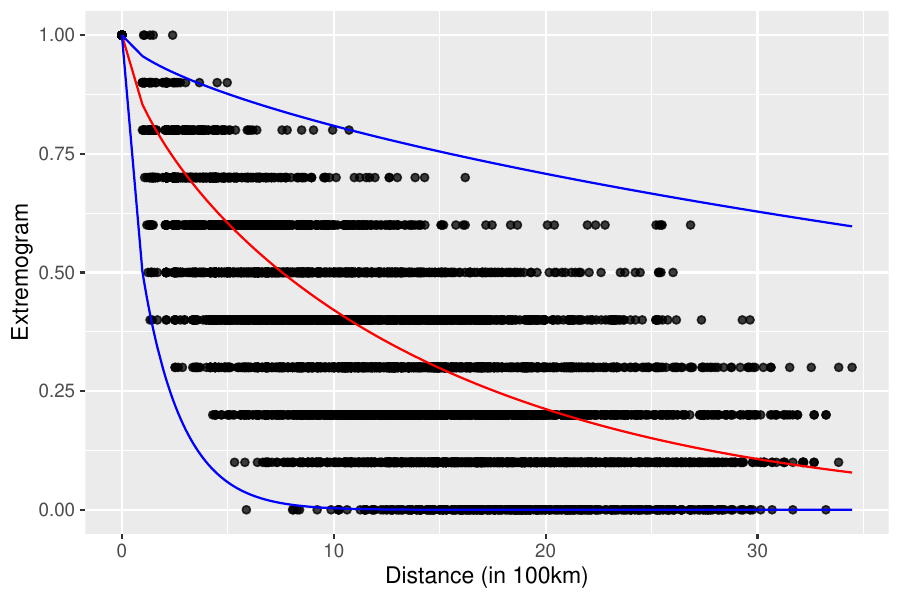}
\end{subfigure}
  \caption{\rev{Empirical estimates of the conditional probability in \eqref{eq:condprob} (with $q=0.75$) for all pairs of grid points from the $r$-exceedances in the test set (dots) varying with distance. The red line shows the average model predicted probability (over all the $r$-exceedance test days), while the blue lines below and above show the ones from the model with artificially low and high dependence in Figure \ref{fig:modelsim}.} }
  \label{fig:extremogram}
\end{figure}

We now compare our proposed spatial sub-model that has spatial extremal dependence varying with predictors, with another that assumes spatial extremal independence \rev{(similar to the simulation with artificially low dependence before)} throughout the study region; the latter uses only the intensity sub-model, and reflects the common practice used in the climate extremes literature when modelling peaks-over-threshold data, e.g., when the GPD is used to model excesses over a high threshold at each location separately. 

Using the average Brier score \citep{Brier.1950}, we evaluate and compare the models' ability to predict joint exceedances above various local empirical quantiles of T2M from the 40 $r$-exceedance days in the test set at two grid points, using the conditional probability \eqref{eq:condprob}, with $q$ the chosen quantile. The severity thresholds are chosen low enough to retain enough observations to evaluate these scores with sufficient precision, but sufficiently high for extreme risk assessment. Note that the thresholds indicated are the grid point-wise empirical quantiles from the $272$ $r$-exceedance days from the training set and not from the full dataset; the quantile of $0.4$ in the grid point containing Paris, for example, corresponds to a high of $+2.2$K. 

Table \ref{table:scores} shows good relative performance of our sub-model throughout. To better grasp the uncertainty in scores, we calculate p-values of a permutation test assessing the differences in scores based on 10'000 permutations. The score differences' 
uncertainty increases for higher thresholds as the score is averaged over fewer observations, though the scores for our spatial model are always better or equal to the other model. This illustrates the benefits of modelling the spatial dependence in the extremes, though some of the high p-values indicate that a validation dataset with a longer observational record would be beneficial, as we are considering only $40$ $r$-exceedance days from the test set. Evaluating models for very extreme events is always challenging, especially with varying spatial dependence; this is an avenue for future research.

\begin{table}[t!]
\centering
\begin{tabular}{ccccc}
  \hline
   & \multicolumn{4}{c}{Q$_{40}$/Q$_{50}$/Q$_{60}$/Q$_{70}$} \\
 Model & London & Paris & Berlin & Amsterdam  \\
 \hline
 Spatial   & 28/25/20/11 & 28/27/26/16 & 27/28/25/26  & 34/34/33/26 \\
 Independent & 34/30/23/11 & 37/35/34/18 & 36/36/32/33 & 39/39/39/29  \\
 p-val & 20/28/38/48 & 14/21/24/41 & 16/22/28/27 &  21/28/30/36\\
  \hline
\end{tabular}
\caption{Mean Brier score calculated from the ERA5 test set (2018-2022) that compares the observed 0-1 indicators for the joint exceedance at grid point $\bm s_1$ and $\bm s_2$ conditional on an $r$-exceedance, against the model-estimated joint probabilities. The empirical quantiles from the $r$-exceedance days are $q=\{0.4,0.5,0.6, 0.7\}$, separated in each cell of the table by ``/''. The grid point $\bm s_1$ corresponds to the grid points containing London, Paris, Berlin or Amsterdam,  and $\bm s_2 $ to a randomly sampled grid point neighbouring $\bm s_1$. P-values ($\times 10^{2}$) are calculated from permutation tests where the alternative hypotheses are that the spatial model does better. A lower score is better. }
\label{table:scores}
\end{table}

\subsection{Code for loss functions and boosting algorithm}\label{sec:code}

Code to run the methods outlined in this paper, along with data files to reproduce the data analyses, are available in the supplementary material of the PRSA article. 

\end{document}